\documentclass{aa}  

\usepackage{graphicx}
\usepackage[colorlinks=true,linkcolor=blue,citecolor=blue]{hyperref}
\usepackage{txfonts}
\usepackage{tabularx}
\usepackage{longtable}
\usepackage{amssymb} 
\usepackage{multirow}
\usepackage{blindtext}
\usepackage{verbatim}
\usepackage{adjustbox}
\usepackage{subfig}
\usepackage{threeparttable}
\usepackage{appendix}
\usepackage{placeins}
\usepackage[normalem]{ulem}
\usepackage{color}
\newcommand{\redcomment}[1]{{\color{red}#1}}

\usepackage{acronym}
\acrodef{VIS}[VIS]{visible}
\acrodef{RV}[RV]{radial velocity}
\acrodef{RVs}[RVs]{radial velocities}
\acrodef{RM}[RM]{Rossiter-McLaughlin}
\acrodef{CLV}[CLV]{centre-to-limb variation}
\acrodef{MCMC}[MCMC]{Markov chain Monte Carlo}
\acrodef{AD}[AD]{absorption depth}
\acrodef{SN}[S/N]{signal-to-noise ratio}

\begin{document} 

\title{The GAPS programme at TNG \\
LXVI. A homogeneous search for \ion{Na}{I} and its possible variability in ten gas giant exoplanets
\thanks{Based on observations made with the Italian Telescopio Nazionale Galileo (TNG) operated on the island of La Palma by the Fundación Galileo Galilei (FGG) of the Istituto Nazionale di Astrofisica (INAF) at the Spanish Observatorio del Roque de los Muchachos of the Instituto de Astrofisica de Canarias. }}

\author{D. Sicilia\inst{1}, L. Malavolta\inst{2,3}, G. Scandariato\inst{1}, L. Fossati\inst{4}, A. F. Lanza\inst{1}, A. S. Bonomo\inst{5}, F. Borsa\inst{6},  G. Guilluy\inst{5}, V. Nascimbeni\inst{3}, L. Pino\inst{7}, F. Biassoni\inst{8,6}, M.C. D’Arpa\inst{9,10}, I. Pagano\inst{1}, A. Sozzetti\inst{5}, M. Stangret\inst{3}, R. Cosentino\inst{11,1}, P. Giacobbe\inst{5}, M. Lodi\inst{11}, J. Maldonado\inst{10}, D. Nardiello\inst{2,3}, M. Pedani\inst{11}}
\institute{
\label{inst:1}INAF – Osservatorio Astrofisico di Catania, Via S. Sofia 78, 95123 Catania, Italy 
\and 
\label{inst:2}Dipartimento di Fisica e Astronomia "Galileo Galilei" - Universit\`a degli Studi di Padova, Vicolo dell'Osservatorio 3, I-35122 Padova, Italy
\and
\label{inst:3}INAF – Osservatorio Astronomico di Padova, Vicolo dell'Osservatorio 5, Padova I-35122, Italy
\and
\label{inst:4}Space Research Institute, Austrian Academy of Sciences, Schmiedl- strasse 6, 8042 Graz, Austria
\and
\label{inst:5}INAF – Osservatorio Astrofisico di Torino, Via Osservatorio 20, 10025, Pino Torinese, Italy
\and
\label{inst:6}INAF – Osservatorio Astronomico di Brera, Via E. Bianchi 46, 23807 Merate, Italy
\and
\label{inst:7}INAF – Osservatorio Astrofisico di Arcetri, Largo E. Fermi 5, 50125, Firenze, Italy
\and
\label{inst:8}DISAT, Universit\`a degli Studi dell’Insubria, via Valleggio 11, I-22100 Como, Italy
\and
\label{inst:9}University of Palermo, Department of Physics and Chemistry “Emilio Segrè, Via Archirafi 36, Palermo, Italy
\and
\label{inst:10}INAF – Osservatorio Astronomico di Palermo, Piazza del Parlamento, 1, I-90134 Palermo, Italy
\and
\label{inst:11}Fundación Galileo Galilei-INAF, Rambla José Ana Fernandez Pérez 7, 38712 Breña Baja, TF, Spain
}
 \date{}
\abstract{
The neutral sodium resonance doublet (\ion{Na}{I} D) has been detected in the upper atmosphere of several close-in gas giants, through high-resolution transmission spectroscopy.
We aim to investigate whether its variability is linked to the planets' properties, the data quality,
or the accuracy of the system parameters used. \\
Using the public code \texttt{SLOPpy}, we extracted the transmission spectrum in the 
\ion{Na}{I} D region of ten gas giants for which a large number of HARPS-N observations are available.
We modelled the absorption signals found, performing an MCMC analysis, and converted the measured absorption depth to the corresponding atmospheric height over which most sodium absorption occurs.\\
While two targets (GJ 436 b and KELT-7 b) 
show no \ion{Na}{I} D feature, we found 
variability in the transmission spectrum of the other 
targets. Three of them (HD 209458 b, WASP-80 b, and WASP-127 b) present 
absorption on only some nights, while in the other five targets (HD 189733 b, KELT-9 b, KELT-20 b, WASP-69 b, and WASP-76 b), a significant absorption signal is present on most of the nights analysed. Except for WASP-69 b, the measured absorption depths 
lead to a ratio 
of the two Na I D depths that is compatible with
or slightly larger than one.
As was expected from literature, the
relative atmospheric height follows an empirical exponential trend as a function of a scaled product of
the planet's equilibrium temperature and surface gravity.\\
We confirm the sodium detection on HD 189733 b, KELT-9 b, KELT-20 b, WASP-69 b, and WASP-76 b.
The signal detected in WASP-127 b requires further observations for definitive confirmation. We exclude a planetary origin for the
signals found on HD 209458 b and WASP-80 b.
The sodium absorption variability 
does not appear to be related to 
planetary properties, but rather to  
data quality, sub-optimal data treatment, or
stellar activity.}

\keywords{planetary systems - planets and satellites: atmospheres - techniques: spectroscopic - planets and satellites: individual: GJ 436 b, HD 189733 b, HD 209458 b, KELT-7 b, KELT-9 b, KELT-20 b, WASP-69 b, WASP-76 b, WASP-80 b, WASP-127 b.}

\titlerunning{A homogeneous search for \ion{Na}{I} and its possible variability}
\authorrunning{D. Sicilia et al.}

\maketitle

\section{Introduction}\label{sec:intro}
To date, more than 5,000 exoplanets belonging to around 4,000 planetary systems have been discovered, and many more are waiting to be confirmed. About one third of the known exoplanets belong to the class of gas giants, similar in mass and size to Jupiter or Saturn in our Solar System.
Over the last two decades, thanks mainly to ground- and space-based transmission spectroscopy, it has been possible to reveal the atmospheric properties of this class of exoplanets, especially those that transit very close to their host star. A wide range of chemical species have been observed with this technique in the terminator of the atmospheres of several hot Jupiters (HJs, e.g. \citealt{Giacobbe_2021, Guilluy_2022}), ultra-HJs (UHJs, e.g. \citealt{Hoeijmakers_2019, Kesseli_2022}), 
and warm Neptunes (e.g. \citealt{Ninan_2020, Basilicata_2024}).
Due to its large absorption cross-section, one of the most investigated atomic species is alkaline-neutral sodium (\ion{Na}{I}, \citealt{Seager&Sasselov}).
In particular, strong signatures in transit spectroscopy occur in the \ion{Na}{I} resonance doublet (\ion{Na}{I} D), characterised by strong lines in the \ac{VIS} range at 
the wavelengths given in the air of 5889.95 \AA \, (D$_2$)
and 5895.92 \AA \, (D$_1$).
Sodium was the first atomic species claimed to be detected in the atmosphere of an exoplanet; namely, HD 209458 b \citep{Charbonneau_2002}. This detection, based on data acquired with the Space Telescope Imaging Spectrograph (STIS) on board the \textit{Hubble} Space Telescope (HST), 
remained one of the most robust examples of atmospheric characterisation, and it has been confirmed by other independent analysis, using the same low-resolution dataset or a different high-resolution one \citep{Sing_2008, Snellen_2008, Albrecht_2009}. 
Later on, using HARPS-N (High Accuracy Radial velocity Planet Searcher for the Northern hemisphere) and CARMENES (Calar Alto high-Resolution search for M dwarfs with Exoearths with Near-infrared and optical Échelle Spectrographs) data, and thanks to improved techniques used for data reduction and increasingly precise observations, \citet{CB_2020} showed that the \ion{Na}{I} signature in the transmission spectrum of HD 209458 b is actually attributable to the combination of the \ac{CLV} and the \ac{RM} effects, which were previously ignored. However, 
\citet{Santos_2020} 
report a tentative broadband detection of \ion{Na}{I} in the planetary atmosphere of this target, obtained using high-resolution data acquired with ESPRESSO (Echelle SPectrograph for Rocky Exoplanets and Stable Spectroscopic Observations, \citealt{ESPRESSO}), mounted on the European Southern Observatory's Very Large Telescope (ESO/VLT). With the same data, \citet{CB_2021} confirmed the non-detection at the \ion{Na}{I} D line cores. This last result seems to confirm that while \ion{Na}{I} wings form at a lower temperature where broadband and low-resolution transmission spectroscopy is sensitive, \ion{Na}{I} cores, sensitive only to higher resolution, form in the upper layers of the atmosphere, where atoms could ionise due to stellar radiation. 

Since the initial ground-based detections of \ion{Na}{I} absorption at high resolution \citep{Redfield_2008, Snellen_2008}, numerous additional detections have been reported over the years for various gas giants. These include observations at low resolution \citep{Nikolov_2013, Chen2017, Chen_2018, Lendl2017, Nikolov_2022} as well as high resolution \citep{Wyttenbach_2015, CB_2017, Seidel_2019, Hoeijmakers_2019, Bello-Arufe_2022}.

In this paper, we present the results of a systematic search for \ion{Na}{I} variability in the transmission spectrum of ten gas-giant planets using high-resolution data retrieved 
using the HARPS-N spectrograph \citep{Cosentino_2012, Cosentino_2014}, mounted on the Telescopio Nazionale Galileo (TNG).
For some targets, we also included archival observations gathered with HARPS-N and HARPS, located in the southern hemisphere at the ESO 3.6-m telescope \citep{Mayor_2003}.

The first aim of this work is to independently verify the detection or non-detection of \ion{Na}{I}, since all the targets in our sample have previously been studied (see Sect. \ref{sec:observations}). Indeed, 
discrepancies exist where detections reported in some studies (not only of sodium but also of other chemical species) are not confirmed when using different datasets (e.g. \citealt{Cauley_2017}) or even the same datasets (e.g. \citealt{Mugnai_2021}). In most cases, the non-homogeneity in the results can be attributed to different assumptions and techniques employed in the analysis.
For this reason, it is important to have public tools (e.g. \citealt{DosSantos2022, Sicilia}) in order to solve the problem of reproducibility for scientific results. However, it should not be excluded that the inconsistent results in the literature could be due to an intrinsic variability of the atmospheric signal itself. This leads us to the main focus of this work: by
studying the \ion{Na}{I} signal variability, we aim to understand how the presence or lack of this atomic species is potentially linked to the planet's properties, the quality of the data, or the accuracy of the system parameters used. 
To achieve this goal, we selected targets with a substantial number of observations: at least five transits for each planet.
By using the same publicly available code, we homogeneously analysed them, and compared the variation in the signals found from night to night.

Furthermore, in the case of detection, we discuss the line depth ratio of the sodium doublet; that is, $f_{D2/D1}$, given by

\begin{equation} \label{eq:line_ratio}
    f_{D2/D1} = \frac{1-\exp{(-\tau_{D2})}}{1-\exp{(-\tau_{D1})}},
\end{equation}

\noindent where $\tau_{D2}$ and $\tau_{D1}$ are the optical depths at the D$_2$ and D$_1$ line centres, respectively. In general, the oscillator strength of the \ion{Na}{I} D$_2$ transition is twice that of the D$_1$ transition, which means that both the absorption cross-section and optical depth at D$_2$ line centre are double the respective values at D$_1$ line centre \citep{Draine_2011}. Assuming that both D lines probe the same region and arise from the same sodium population, \citet{Gebek-Oza} performed both forward and inverse modelling of high-resolution transmission spectra in the sodium doublet for different scenarios 
(i.e. hydrostatic, escaping, exomoon, and torus).
What they found is that $f_{D2/D1}$ 
approaches two 
in the optically thin regime ($\tau < 0.1$), while it decreases to one in the optically thick regime ($\tau > 10$). Recent studies on the measurement of $f_{D2/D1}$ seem to have confirmed this \citep{ Zak, Khalafinejad_2021}. However, the authors of the work note that, from the measurement of this ratio, it is only possible to deduce in which regime the majority of the absorption occurs, which does not necessarily correspond to the regime throughout the area of the stellar disc blocked by the planet. Besides, the line strength and the mixing ratio depend on the composition of the surrounding atmosphere, which can even vary over time. As an example, \citet{Slanger_2005} showed that, in the Earth’s atmosphere, $f_{D2/D1}$ is not constant but varies between 1.2 and 1.8. 

\citet{Langeveld_2022}, hereafter L2022, presented the first homogeneous survey of sodium absorption in ten highly irradiated giant exoplanets using high-resolution transmission spectroscopy retrieved with HARPS and HARPS-N. They confirmed nine previous detections and reported a new detection with a single transit. Measuring $f_{D2/D1}$ across their sample, they found seven targets with line ratios consistent with unity, and three with line ratios greater than 1. However, due to the size of their uncertainties, they do not make a conclusive statement about the line ratios, but suggest combining further observations. In this first survey, the importance of the quality of the data is evident (e.g. a weighted average is more appropriate than a simple average when combining all spectra), as is the position of the planetary features within the residual spectra (e.g. the spurious signals induced by \ac{CLV} and the \ac{RM} effect
may overlap with a potential planetary absorption).

This paper is organised as follows. In Sect. \ref{sec:observations}, we present the sample and the observations used in this work. The methods applied for the data reduction and the analysis are described in Sect. \ref{sec:analysis}.
In Sect. \ref{sec:results} and Sect. \ref{sec:interpretation}, we discuss and interpret our results, respectively. Finally, in Sect. \ref{sec:conclusions} we summarise our findings and provide our conclusions.

\section{The sample and the observations}\label{sec:observations}

This work is part of the Global Architecture of Planetary Systems (GAPS) programme \citep{GAPS_Covino}.
GAPS is a long-term, Italian project that investigates several open issues in the field of exoplanetary science thanks to the availability of the HARPS-N spectrograph \citep{Cosentino_2012, Cosentino_2014} located at the Italian telescope TNG in La Palma (Canary Islands, Spain). 

One of the main lines of research of GAPS is focussed on the study and characterisation of a number of HJ atmospheres, both in the \ac{VIS} (e.g. \citealt{Borsa_2022}) and near infrared (nIR) wavelength range (e.g. \citealt{Carleo_2022}). Within GAPS, we have collected various transit observations and phase curves for more than 20 targets which have led to the publication of several works (e.g. \citealt{Esposito_2014, Mancini_2015, Bonomo_2017, Mancini_2018, Borsa_2019, Guilluy_2020, Scandariato_2021, Pino_2022, Fossati_2023, Sicilia_hatp67}). The programme on atmospheres found its natural continuation in BRIDGES (Building a Road to the In-Depth investiGation of Exoplanetary atmosphereS, P.I. Borsa), which aims to perform a deep, time-intensive, simultaneous \ac{VIS} and nIR characterisation of a selected sample of exoplanetary atmospheres.

\begin{figure}
    \centering
    \includegraphics[width=\linewidth]{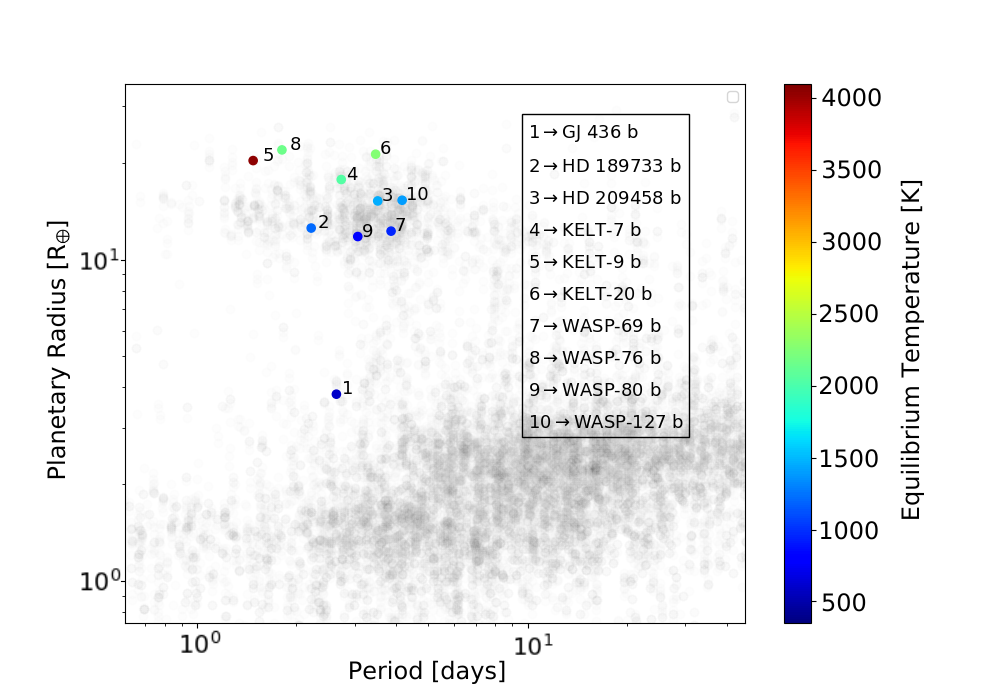}
    \caption{Known exoplanets as a function of their radius and period. The targets analysed in this work are highlighted. The marker colour reflects their equilibrium temperature. Data taken from the NASA Exoplanet Archive \citep{nasa-archive}.}
    \label{fig:sample}
\end{figure}

Here, we report on the analysis in the \ac{VIS} range around the \ion{Na}{I} D (5870 - 5915 $\AA$) 
of ten gas giants: GJ 436 b, HD 189733 b, HD 209458 b, KELT-7 b, KELT-9 b, KELT-20 b, WASP-69 b, WASP-76 b, WASP-80 b, and WASP-127 b (see Fig. \ref{fig:sample}).
\href{https://zenodo.org/records/14268745}{Table B.1} summarises the observations of each exoplanet transit, while Tables \ref{tab:pams1} and \ref{tab:pams2} show a compiled list of the main stellar and planetary 
parameters that are used in the analysis. 

In the following, we give a brief description of each target analysed in this work along with an overview of the observations used. 

\subsection{GJ 436 b} GJ 436 b \citep{Butler_2004}, also known as Gliese 436 b, is a warm Neptune-mass (M$\rm_p$ = 0.07 M$\rm_J$) planet. It orbits an M2.5 star (V=10.68) with a period of $\sim$ 2.6 days.

For this target, five transits in the \ac{VIS} were gathered in the framework of the GAPS programme, and two transits under the BRIDGES programme, using low-cadence (900 s) exposures. This resulted in a total of 84 spectra. We discarded the last spectrum of the second night (GJ-N2), the acquisition of which was stopped only after about 3 seconds, and one in-transit spectrum of the third night (GJ-N3) because it was leading to a crash in the telluric correction phase. 
The last two nights (GJ-N6 and GJ-N7) were characterised by very variable seeing.

\subsection{HD 189733 b} \label{sec:hd189}
HD 189733 b \citep{Bouchy_2005} is among the best-studied planets to date. It is an HJ with approximately the same mass and radius as Jupiter, 
orbiting a visual binary bright (V $\sim 7.7$) and active K-type star every $\sim 2.2$ days.

For this target, we analysed four transit events observed as part of the GAPS programme and combined them with other four transits retrieved with HARPS-N (programmes: A35TAC\_14 
and BRIDGES), and three transits retrieved with HARPS (ESO programmes: 072.C-0488(E), 079.C-0127(A)
). 
The second HARPS night (HD1-N2) has no observations before transit and an interruption of 20 minutes after the egress; the third HARPS night (HD1-N3) presents an interruption of 17 minutes after the transit; the second HARPS-N night (HD1-N5) has an interruption of 47 minutes after the transit (elevation too low, target under the horizon); the fourth HARPS-N night (HD1-N7) has an interruption of 25 minutes after the transit; the fifth HARPS-N night (HD1-N8) is characterised by a rapid increase in relative humidity; in the last GAPS night (HD1-N9), we discarded the last spectrum due to its low \ac{SN} compared with the rest of the dataset. 
In the end, our analysis for this target is based on 366 spectra covering eleven planetary transits.
In these observations, sky spectra were gathered only for HD1-N2, HD1-N3, and from HD1-N8 to HD1-N11, while on other nights fibre B pointed to a Fabry-Perot lamp. We applied sky correction 
when possible.\\

\subsection{HD 209458 b} \label{sec:hd209}
HD 209458 b \citep{Henry_2000} is one of the most favourable targets for atmospheric characterisation; indeed, several species have been discovered in its atmosphere (e.g. \citealt{Giacobbe_2021}).
The planet orbits a bright (V=7.65) sun-like (G0V) star in a 3.5-days period.

For this target, we analysed three transits retrieved with HARPS-N which are publicly available in the TNG archive (programmes: A32TAC$\_$41 and A35TAC$\_$14) and five GAPS transits. Two other public transits were available but they have not been taken into account because the \ac{SN} of the observations was too low. During the ingress of the first night (HD2-N1) there was an interruption of about 36 minutes in the observations due to the humidity being too high; during the egress of the fourth night (HD2-N4) there was an interruption of almost 1 hour due to 
technical problems; the sixth night (HD2-N6) was characterised by multiple stops due to clouds. In the end, our analysis is based on 243 spectra covering eight planetary transits. 
The sky correction was skipped for HD2-N2 and HD2-N3 because fibre B pointed to the Fabry-Perot lamp instead of the sky.

\subsection{KELT-7 b} \label{sec:kelt7}
KELT-7 b \citep{Bieryla_2015} is an UHJ transiting a rapidly rotating ($v \sin{i} = 71.4 \pm 0.2 $ km$\, $s$^{-1}$) F-star every $\sim$ 2.7 days.
Given its relatively low surface gravity ($\log g\rm_p$ $\sim$ 3.1), high equilibrium temperature (T$\rm_{eq}$ $\sim$ 2048.0 K), and bright (V=8.54) host star, KELT-7 b is an ideal target for detailed characterisation.

For this target, there are four transits retrieved during the GAPS programme and one public transit available from the TNG archive (programme: CAT17B\_29).
This last night (K7-N3) covers only the pre-transit and half-transit phase while, both the first and fourth GAPS nights (K7-N1 and K7-N5) are characterised by a stop during the transit, almost 2 hours and 30 minutes, respectively. Besides, one low-flux spectrum was discarded both on K7-N3 and K7-N5.
Nevertheless, for this target, we have been able to analyse a total of 131 spectra, of which 72 in-transit. 
Only K7-N1 does not have the necessary 
files to perform the sky correction, which has therefore been neglected.

\subsection{KELT-9 b} \label{sec:kelt9}
KELT-9 b \citep{Gaudi_2017} is an UHJ 
orbiting a rapidly rotating ($v \sin{i} = 111.40 \pm 1.27 $ km$\, $s$^{-1}$), oblate, early A-type star every $\sim$ 1.5 days. 
The host star is more than twice the radius of the Sun and has an effective temperature of roughly 10,000 K, making it $\sim$ 50 times more luminous. At any given time, KELT-9 b receives $\sim$  44,000 times as much incident flux as Earth.
Due to this intense stellar irradiation and
to the close orbital distance (0.03 AU), KELT-9 b is 
the hottest transiting planet known
to date (T$\rm_{eq} \sim$ 4000 K). 

For this target, we analysed four GAPS transits 
and combined them with two transit events (programmes: A35DDT4 and OPT18A-38) 
retrieved again with HARPS-N.
The second and third GAPS nights (K9-N4 and K9-N5) are characterised by a variable seeing and a short pre-transit phase, respectively. For K9-N5, we had to remove one in-transit spectrum from the analysis due to its very low \ac{SN} compared to the rest of the dataset (see \href{https://zenodo.org/records/14268745}{Fig. B.1}). In the end, our analysis is based on 320 spectra covering six planetary transits.\\ 

\subsection{KELT-20 b} \label{sec:kelt20}
KELT-20 b \citep{Lund_2017}, also named MASCARA-2 b \citep{Talens_2018}, is one of the most studied UHJs (T$\rm_{eq} \sim $ 2260 K). 
It transits a rapidly rotating ($v \sin{i} = 115.9 \pm 3.4 $ km$\, $s$^{-1}$) A-type star with an orbital period of $\sim$ 3.5 days. 

For this target, we analysed three GAPS transits 
and combined them with other three public HARPS-N transits 
(TNG archive, programmes: CAT17A$\_$38 and CAT18A$\_$34). 
We discarded eight consecutive out-of-transit spectra in the second night (K20-N2) which presented a lower \ac{SN} due to clouds and one in-transit spectrum which also presented a similar low \ac{SN}. For the same reason, we also removed three consecutive in-transit spectra in the last GAPS night (K20-N6). In the end, our analysis of KELT-20 b is based on six planetary transits consisting of 361 spectra.\\ 

\subsection{WASP-69 b} \label{sec:wasp69}
WASP-69 b \citep{Anderson_2014} is an inflated Saturn-mass planet ($\sim$0.26 M$\rm_J$). It orbits
an active K-type star (9.87 V) with a period of $\sim$ 3.9 days. 
 
For this target, we analysed a total of 131 spectra covering six planetary transits: three of them retrieved as part of the GAPS programme, 
two public transits available from the TNG archive (programme: CAT16A\_130), and the last one retrieved in the framework of the BRIDGES programme. 
In the framework of the GAPS programme, WASP-69 b was also observed during three other nights (2021-10-28, 2022-09-14, and 2022-10-15) but, due to misalignment errors, the observations were completed with GIANO only. 
As shown in \href{https://zenodo.org/records/14268745}{Fig. B.1} there are two spectra with a higher noise compared to other spectra of the same night: the last spectrum of the night 2016-06-03 (W69-N1) and the first spectrum of the night 2020-08-09 (W69-N4). We excluded both spectra. 

\subsection{WASP-76 b} \label{sec:wasp76}
WASP-76 b \citep{West_2016} is one of the most inflated UHJs (T$\rm_{eq} \sim $ 2160 K) studied to date. It orbits an F-type star every $\sim$ 1.8 days.

In this work, we analysed five transit events from the GAPS programme and combined them with three transits available from public HARPS data (ESO programmes: 100.C-0750; 
090.C-0540)
and one transit retrieved with HARPS-N in the framework of the BRIDGES programme.
In the end, our analysis is based on 297 spectra covering nine planetary transits. 
In the third HARPS night (W76-N4), all data after planet transit were excluded from the analysis due to clouds, so the master-out of the specified night was created using just the spectra taken before the transit. The radial velocities of the first HARPS night (W76-N1) present an outlier at the beginning of the night, with a similar \ac{SN} with respect to the other observations of the spectroscopic series, and no evident correlation with the weather condition in La Silla during that night \citep{Seidel_2019}. As a precautionary measure, this observation was excluded from the analysis. Sky correction was applied for all nights except for the first one because sky spectra on fibre B were not available.

\subsection{WASP-80 b} \label{wasp80}
WASP-80 b \citep{Triaud_2013} is an HJ orbiting a rather active K-type star every $\sim$ 3.07 days. 
Its low density ($\sim$0.8 g cm$^{-3}$) and the large transit depth ($\sim$3\%) make it a very good candidate for transmission spectroscopy.

For this target, eight transits were gathered in the framework of the GAPS programme. However, we decided to completely discard one transit from the analysis (2020-09-17) because it was characterised by excessive telluric sodium emission and a lower \ac{SN} compared to the other nights, probably due to the presence of thin clouds. In addition, we removed the first low-\ac{SN} four spectra (three out-of-transit, one in-transit) collected during W80-N4. In the end, our analysis is based on 104 spectra covering seven planetary transits.

\subsection{WASP-127 b} \label{sec:wasp127}
WASP-127 b \citep{Lam_2017} is a heavily bloated gas exoplanet with one of the lowest densities discovered to date ($\sim 0.09$ g cm$^{-3}$). 
It orbits a bright G5 star (V=10.15) every $\sim$ 4 days. 

For this target, we analysed three GAPS transits and publicly available HARPS data, covering three transit events as part of the HEARTS survey (ESO programme: 098.C-0304, 0100.C-0750(D)).
Of the HARPS nights, 
the first one (W127-N1) has no observations during the egress and after the transit while the second and the third night (W127-N2 and W127-N3) have a very short post-transit phase. This is due to the visibility constraints of the target, especially its long transit duration (> 4 hours). We removed the last spectrum of W127-N1 for low \ac{SN} and that one of W127-N3 because the airmass was too high ($\sim 2.7$).
Of the GAPS nights, the first one (W127-N4) is contaminated by moonlight, and the second one (W127-N5) is characterised by a very variable seeing. We decided to analyse both of them anyway but excluding from the analysis the first three out-of-transit spectra of W127-N1 that left a strong residual in the final transmission spectrum.

\section{Analysis} \label{sec:analysis}
\subsection{Data reduction} \label{sec:data_reduction}
The standard data reduction of the raw spectra, including bias subtraction, flat-fielding and wavelength calibration is already performed by the HARPS and HARPS-N Data Reduction Software (DRS) which produces both 2D spectra (referred to the observer reference frame) and 1D spectra (referred to the barycentre of the Solar System). For each night of observation, we analysed the 2D spectra using the \texttt{SLOPpy}\footnote{\href{https://github.com/LucaMalavolta/SLOPpy}{https://github.com/LucaMalavolta/SLOPpy}} (Spectral Lines Of Planets with python) public pipeline \citep{Sicilia}. 

\texttt{SLOPpy} is a user-friendly, standard and reliable tool which is optimised for the spectral reduction and the extraction of transmission planetary spectra in the \ac{VIS} obtained from high-resolution observations. For this purpose, \texttt{SLOPpy} first applies several data reduction steps that are required to correct the input spectra for: sky emission (being corrected only when the sky source is retrieved simultaneously with the target thanks to a dedicated fibre, named fibre B), atmospheric dispersion, and presence of telluric features. 
For a detailed description of each individual step, see \citet{Sicilia}.

During the HARPS-N nights acquired between 2017-10-26 and 2018-07-23, a malfunctioning Atmospheric Dispersion Corrector of the telescope introduced strong variations on the overall spectral energy distribution and a decrease in the \ac{SN} (e.g. see \citealt{Borsa_2019}). However, we decided to keep these nights in the analysis since the differential refraction effect is properly corrected by \texttt{SLOPpy} and it did not affect the final transmission spectrum, as we verified.

Next, the pipeline averages all the out-of-transit spectra (i.e. those acquired before the ingress and after the planet's egress) to build the master-out spectrum ($M_{\rm OUT}$) for each night. In general, the out-of-transit spectra are first aligned to the same stellar rest frame (SRF) by using the stellar \ac{RV} computed from the orbital solution of the system ($K_\star$). Nonetheless, for fast rotators of our sample (i.e. KELT-7, KELT-9 and KELT-20, $v\sin{i} \geq$ 60 km$\, $s$^{-1}$), the spectral line broadening is much larger than the \ac{RV} shift induced by the presence of an HJ (as a reference, an 8 M$\rm_J$ would induce keplerian amplitude of the order of 1 km$\, $s$^{-1}$, approximately corresponding to one HARPS/HARPS-N pixel). This makes the correction of the keplerian wobble of the star negligible and, for this reason, we decided to artificially set $K_\star$ = 0 km$\, $s$^{-1}$. This approach also has the advantage that the interstellar absorption lines present in their spectra are at the same wavelength in the whole spectral series and are not diluted when the average out-of-transit $M_{\rm OUT}$ is computed. For the other targets in our sample, closer to the Earth, no interstellar sodium is observed.

\texttt{SLOPpy} then divides each in-transit spectrum (which should contain the planetary signal) by the $M_{\rm OUT}$, in order to extract the transmission spectrum (see Sect. \ref{sec:extraction_ts}). The division also automatically removes the interstellar absorption lines (see e.g. \citealt{CB_2018}). 
Although we did not explicitly analyse the behaviour of the interstellar lines throughout the night,
we note that any change in their shape or position in the Solar System Barycentric reference frame would have caused a smear in the average interstellar lines, resulting in a systematic residual 
when dividing by the $M_{\rm OUT}$.
As a safety check, we routinely produced transmission spectra in the SRF, where no systematic residuals were found.
If a variation in the interstellar lines were present, it was below the noise level of each observation. 

In some cases, after telluric correction performed with version 3.8 of \texttt{Molecfit} \citep{Smette_2015, Molecfit_Kausch}, which is wrapped within \texttt{SLOPpy} through a system call, some residuals are still present. In these cases (i.e. HD 209458 b, KELT-9 b and KELT-20 b), when dividing each spectrum by the $M_{\rm OUT}$, we 
replaced the remaining telluric residuals  
with a linear spline fit of the remaining spectra.


\subsection{Extraction of the transmission spectra}\label{sec:extraction_ts}
After the data reduction, for each target, and for each transit, we extracted the final transmission spectrum $\tilde{\mathfrak{R}}$, given by the weighted average of all the spectral residuals, obtained by dividing each fully in-transit observation ($F_{i,in}$) by the master-out normalised to unity ($\tilde{M_{\rm OUT}}$):

\begin{equation}\label{eq:transmission spectrum}
\tilde{\mathfrak{R_i}} = \frac{F_{i \rm,in}}{\tilde{M_{\rm OUT}}}.
\end{equation}

To take into account the planet's motion around its star, the calculation of $\tilde{\mathfrak{R}}$ is performed by \texttt{SLOPpy} in the planetary reference frame (PRF), by using the theoretical value of the \ac{RV} semi-amplitude of the planet ($K_p$). This avoids the smearing of the planetary atmospheric absorption lines that we would obtain by performing this step in the stellar or observer reference frame \citep{Wyttenbach_2015}.

Before combining all the spectral residuals, the pipeline corrects them for \ac{CLV} and \ac{RM} effects, the two main processes altering the transmission spectra. 
The \ac{CLV} effect refers to the change in the profile of the normalised stellar lines when moving from the centre of a star's disc to the limb. This variation occurs because the physical conditions (such as temperature, pressure, and density) and the line of sight angle in the line formation layer change across the stellar disc. The \ac{RM} effect is a consequence of the rotation of the star. The stellar light blocked by the transiting planet may be redshifted or blueshifted, depending on which side of the star the planet is covering, thus causing a deformation of each spectral line.
The stellar models for each target used for the correction of these effects
have been obtained with Spectroscopy Made Easy (SME, \citealt{Piskunov-Valenti}) using the line list from the VALD database \citep{Ryabchikova_2015} and MARCS \citep{Gustafsoon_2008} or Kurucz ATLAS9 \citep{Kurucz_ATLAS} grids. 
The \ac{CLV} and \ac{RM} modelling includes a factor $r$ to take into account the possible difference in the planetary radius (R$\rm_p$) in the wavelength range under analysis (around the sodium doublet) with respect to the value obtained with transit photometry (likely gathered in a different wavelength range). To make the analysis as homogeneous as possible, the correction of the \ac{CLV} and \ac{RM} effects is applied to all ten targets in the sample, even in those where these effects were negligible (e.g. GJ-436 b, WASP-76 b).

\subsection{Markov chain Monte Carlo analysis}\label{sec:mcmc_analysis}
If there are hints of an absorption signal in $\tilde{\mathfrak{R}}$,
we tried to model it and estimate the detection significance, performing a \ac{MCMC} analysis with the \texttt{emcee} tool \citep{MCMC}. The model assumes a double Gaussian profile for the \ion{Na}{I} absorption lines, and a flat spectrum ($\tilde{\mathfrak{R_i}}$ = 1) otherwise. 
The free parameters of the model are: 
$K_p$, required to model the atmospheric absorption lines in the PRF; the contrasts ($c$ D$_2$, $c$ D$_1$), and the full width at half maximum (FWHM) of the double Gaussian profile describing the absorption signal; the \ac{RV} of the atmospheric wind ($v_{wind}$), which is the shift relative to the PRF transition; the effective planet radius scale factor ($r$). We assumed the same FWHM and $v_{wind}$ for both D lines. To take into account any additional systematics (e.g. bad seeing), we also allow for a free jitter parameter in the fitting procedure. The median values of the posteriors 
are adopted as the best-fit values, and their error bars correspond to the 1$\sigma$ statistical errors at the corresponding percentiles. 

The analysis was performed on the individual spectral residuals $\tilde{\mathfrak{R_i}}$ in the SRF; that is, we computed the model in the 2D phase-wavelength space and then the model was fit to the spectral residuals for each observed phase.
For each sampling step of the \ac{MCMC}, we first applied the \ac{CLV} and \ac{RM} correction corresponding to the sampling value of $r$. Then, we computed the atmospheric absorption signal in the PRF, we applied the Doppler shift due to $v_{wind}$, and finally we shifted the model in the SRF according to the velocity of the planet orbital phase at the time of the observation. This operation was repeated for each $\tilde{\mathfrak{R_i}}$ before computing the log-likelihood. By preserving both temporal and wavelength information, rather than comparing the model with the average final transmission spectrum $\tilde{\mathfrak{R}}$, we are avoiding degeneracies between time-dependent effects such as the velocity of the planet (driven by $K_p$), and wavelength-only effects such as the atmospheric winds on its surface. 
When the \ac{MCMC} fitting is performed, the average final transmission spectrum $\tilde{\mathfrak{R}}$ is re-calculated using the median value of the resulting posterior for the parameters included in the fit. More details can be found in \citet{Sicilia}. We note that the \ac{MCMC} analysis was applied only to fully in-transit data (i.e. between the 2nd and 3rd contact time period), since the average absorption signal differs significantly during ingress and egress (e.g. see \citealt{keles2024pepsi}).

When the \ac{MCMC} analysis returns a significant detection (>3$\sigma$), we also calculate the \ac{AD} of the signal by integrating the flux across narrow passbands centred on the \ion{Na}{i} doublet (one for each line, being then the \ac{AD} of the two lines averaged), and by comparing it with the flux integrated into a reference passband in the continuum. 
For the central bandwidth we use: 2 $\times$ 0.75 \AA, 2 $\times$ 1.50 \AA, 2 $\times$ 3.00 \AA; for the reference passbands in the continuum, we use $[5874.89 - 5886.89]$ \AA \, for the blue side and $[5898.89 - 5910.89]$ \AA) for the red side. 
We extracted the \ac{AD} following two different approaches: analysing 
$\tilde{\mathfrak{R}}$ \citep{Redfield_2008}, or analysing the transmission light curve (TLC), which is the relative flux in a specific passband as a function of time \citep{Charbonneau_2002, Snellen_2008} (for more details, see Sect. 3 in \citealt{Sicilia}).

\begin{figure*}
    \centering
    \includegraphics[width=\textwidth]{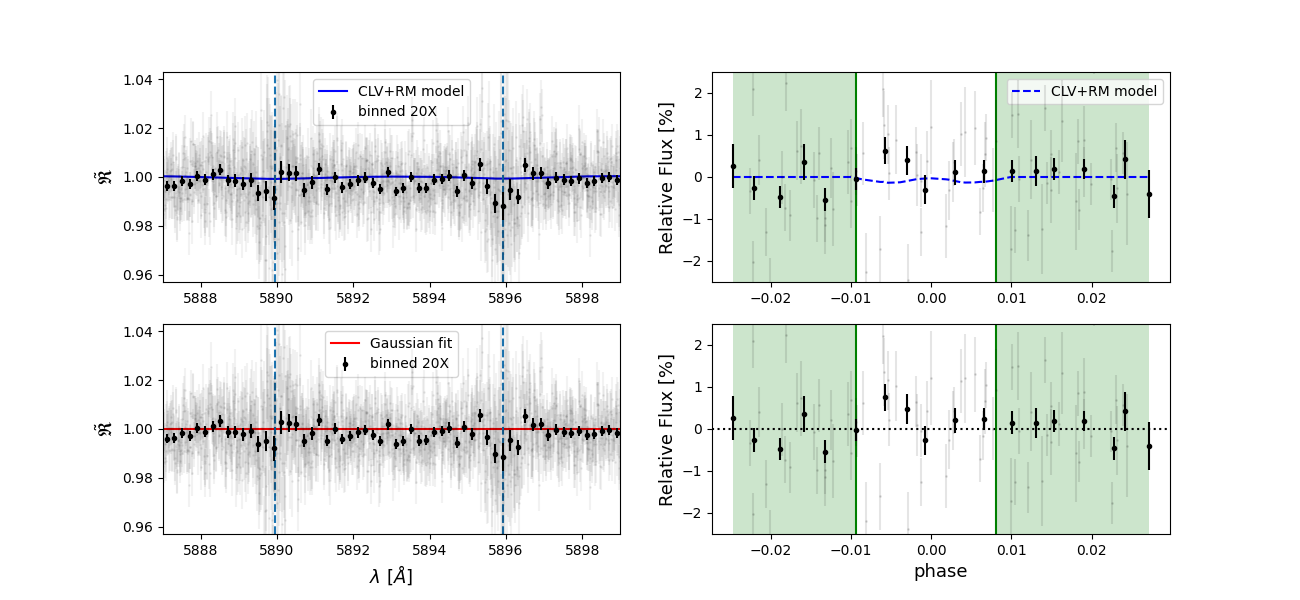}
         \caption{Final transmission spectrum (\textit{left}) and relative TLC (\textit{right}) of GJ 436 b without the correction of the \ac{CLV} and \ac{RM} effects (\textit{upper panels}) and after their correction (\textit{bottom panels}). The green background marks the exposures taken out-of-transit. For graphic purposes only, we fixed the FWHM of the Gaussian fit (red line) at zero being equal to 0.94$^{+0.41}_{-0.31}$ km$\,$s$^{-1}$.}
         \label{fig:GJ436}
\end{figure*}
\begin{figure*}
    \centering
    \includegraphics[width=\textwidth]{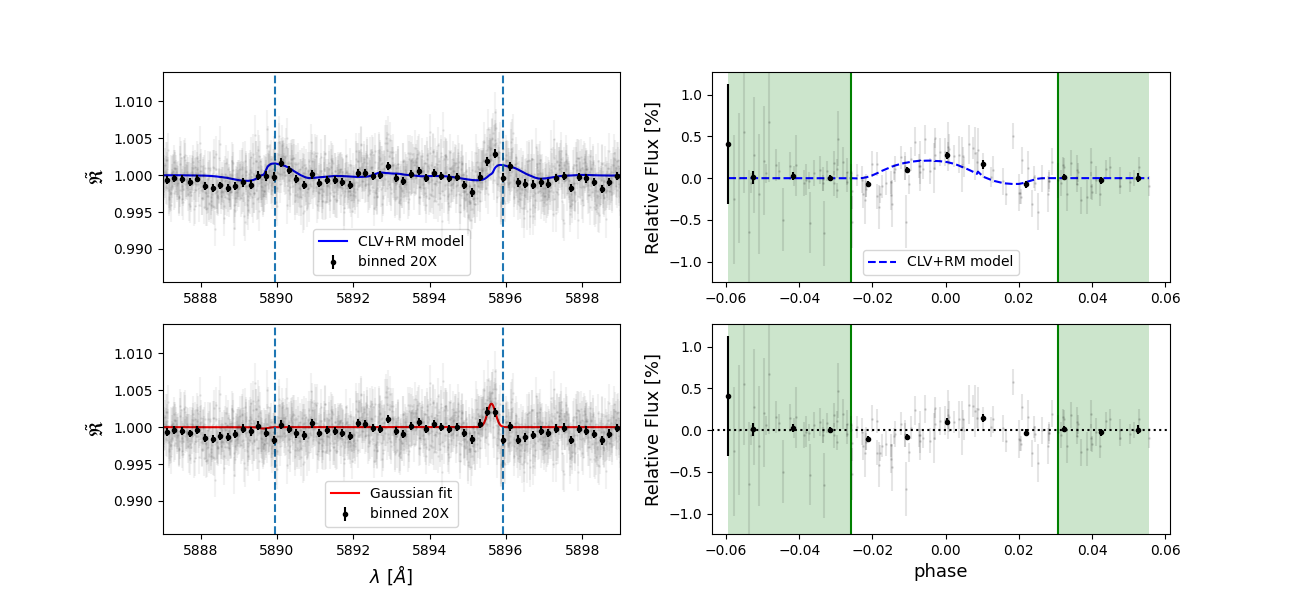}
    \caption{Same as Fig. \ref{fig:GJ436} but for KELT-7 b. For graphic purposes only, we fixed the D$_2$ line contrast of the Gaussian fit at zero being equal to -0.02$^{+0.09}_{-0.11}$.}
    \label{fig:Kelt7}
\end{figure*}

\section{Results}\label{sec:results}
In this section we present the results obtained for each target, distinguishing targets without detection (i.e. GJ 436 b, KELT-7 b), targets in which sodium is significantly detected (i.e. HD 189733 b, KELT-9 b, KELT-20 b, WASP-69 b and WASP-76 b), and targets with no clear detection (i.e. HD 209458 b, WASP-80 b and WASP-127 b). 

\subsection{Targets without detection}\label{sec:no-detections}
For two targets in the sample, namely GJ 436 b and KELT-7 b, the \ac{MCMC} fit did not converge on any significant signal in absorption. 

\begin{itemize}

     \item GJ 436 b. As is shown in Fig. \ref{fig:GJ436}, $\tilde{\mathfrak{R}}$ and the final TLC of this target are almost flat and do not seem to show signs of planetary absorption. This is true for each individual night, where the \ac{MCMC} analysis converges on some emission features, rather than absorption ones. However, these features in correspondence of the \ion{Na}{I} D lines, are unlikely to be of astrophysical origin but could be due to a low \ac{SN} systematic residual. This would also explain the absorption feature visible in 
     $\tilde{\mathfrak{R}}$
     at line D$_1$, which is in any case not fitted by the \ac{MCMC} analysis.
     As was already expected, the correction for the \ac{CLV} and \ac{RM} effects could be neglected for this target. Indeed, the host star is a slow rotator ($v\sin{i}$ = 0.330$^{+0.091}_{-0.066}$ km$\, $s$^{-1}$) and the eccentric orbit of GJ 436 b is nearly perpendicular to the stellar equator \citep{Bourrier_2018}, so the \ac{RM} effect is not visible, as well as the change in the depth of the line core. \\
     It must be noted that the long exposure time, which reduces the number of in-transit spectra (see \href{https://zenodo.org/records/14268745}{Table B.1}), 
     can result in a broadening of the absorption features,
     thus smearing the planetary signal over multiple detector pixels \citep{Cauley_2021, Boldt-Christmas2024}.
     However, we verified that the smearing effect deriving from a single acquisition ($\sim$ 3.7 pixels) can be neglected in this case.
     On the other hand, from the calculation of the planetary velocity during the entire transit (varying from $\sim$ -7 to +7 km$\, $s$^{-1}$), we checked that 
     a possible atmospheric signal would be Doppler shifted of about 0.14 \AA, overlapping with the low \ac{SN} residuals from the stellar line cores. 
     As a consequence, our ability to detect a planetary signal is severely compromised.  
     Shorter exposures, and possibly at higher \ac{SN}, would make the detection less challenging.\\    
     Our results are in agreement with \citet{Knutson_2014} who reported an effectively featureless transmission spectrum using the red grism on the HST/WFC3 (Wide Field Camera 3) instrument,  \citet{Lanotte_2014} who combined Spitzer photometry with HARPS spectroscopy, and \citet{Lothringer_2018} who found no evidence for alkali absorption features using HST/STIS data.
     While the absence of features 
     is compatible with either high altitude obscuring clouds or high metallicity 
     atmosphere models, a large Ly $\alpha$ absorption feature, found by \citet{Ehrenreich_2015}, indicates that the planet is surrounded and trailed by a large exospheric cloud composed mainly of hydrogen atoms, due to energetic stellar irradiation. 
     \\
    \item KELT-7 b. In this target, the \ac{RM} effect (that has an unusually large amplitude of several hundred m s$^{-1}$, \citealt{Bieryla_2015}) is clearly visible (Fig. \ref{fig:Kelt7}).  
    After removing the contribution of the \ac{CLV} and \ac{RM} effects (bottom panel), no clear feature is visible above the noise level, 
    except for an emission residual at the D$_1$ line. 
    We verified that this feature comes from K7-N1; as a matter of fact, it disappears if we exclude this night from the analysis. Its origin is likely telluric since that we could not perform the sky correction for this night (see Sect. \ref{sec:kelt7}).
   Examining individual nights, we found a flat transmission spectrum except on K7-N1 and K7-N5, where the \ac{MCMC} analysis fit a double absorption in correspondence to the sodium lines; however, these absorption features are compatible with scatter.  
   We emphasise that we would not be able to investigate a possible time-dependent variation throughout the transit (e.g. \citealt{Gandhi_2023, Prinoth_2023}) due to the significant amount of spectra missing during three of the five nights analysed (see Sect. \ref{sec:kelt7}),
 even in the case of a significant detection.
    \\
   Our results are in agreement with \citet{Stangret-Kelt7}, who analysed K7-N2 and K7-N3,
   and did not detect any atmospheric species, and with \citet{Tabernero_Kelt7}, who were only able to determine an upper limit of 0.076 \% for the \ion{Na}{I} D, using HORuS (High Optical Resolution Spectrograph) data.
   Being a favourable target for transmission spectroscopy, the lack of absorption features in KELT-7 b could be due to the presence of clouds or hazes in its atmosphere, as well as the presence of the strong \ac{RM} effect or stellar pulsations that can cover or mimic the planetary signal at high resolution. 
   Another hypothesis, which could also be valid for GJ 436 b, is that the noise level is too high for a detection to be visible. We refer the reader to Sect. \ref{sec:interpretation} for a more in-depth discussion. 
 \end{itemize}

\subsection{Targets with detection}\label{sec:detections}
\begin{figure*}
    \centering
     \includegraphics[width = \textwidth]{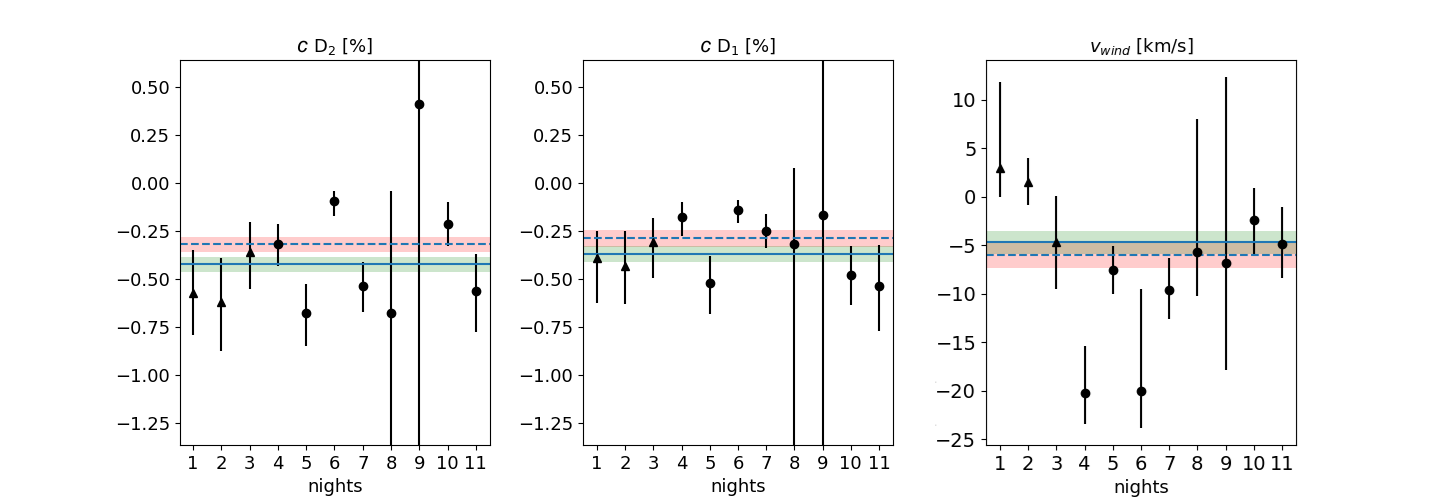}\\
    \includegraphics[width=\textwidth]{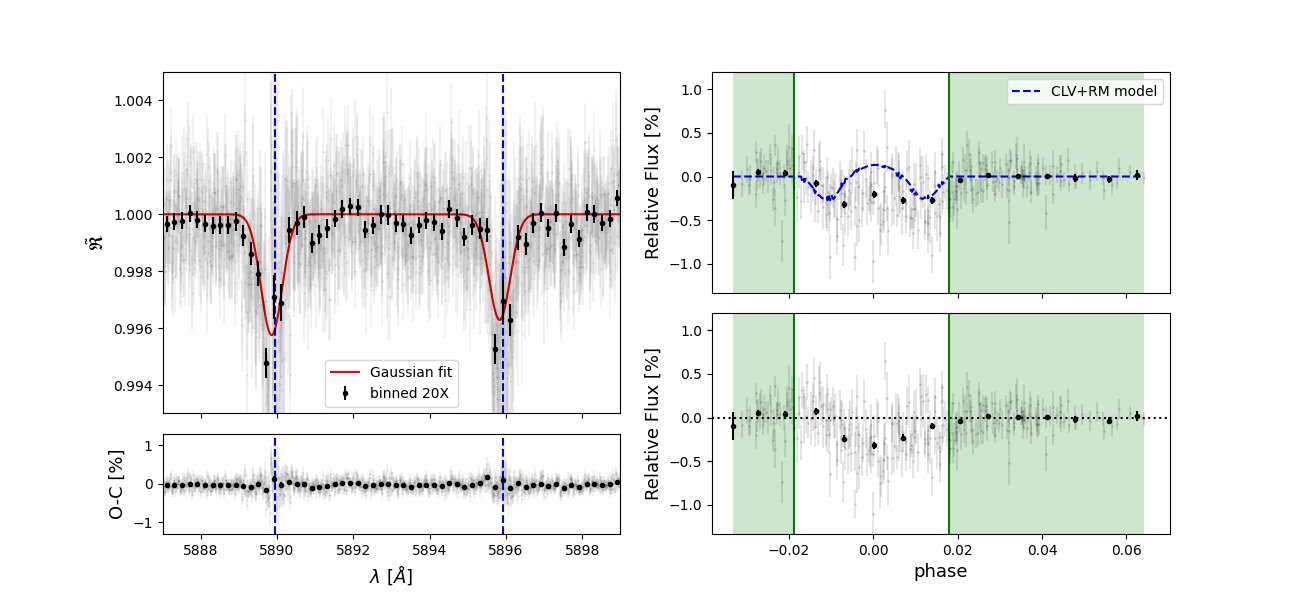}
    \caption{Results obtained for HD 189733 b. \textit{Upper panels:} Contrasts of Na D$_2$ and D$_1$ lines (first and second panel) and 
    wind measurements (third panel) for each night.
    Data points marked with triangles correspond to HARPS nights, circles to HARPS-N nights. The value obtained combining all nights is indicated by a horizontal dashed line, while its 1$\sigma$ errors mark the confidence interval, which is indicated with a red band. The horizontal solid line inside the green band indicates the average value obtained after the exclusion of the nights with larger uncertainties or very small contrast (see the text). 
   \textit{Bottom panels:} Final transmission spectrum 
   centred around the \ion{Na}{I} D  (light grey), also binned by 20x in black circles (\textit{left}) and the relative TLC, before and after the correction of the \ac{CLV} and \ac{RM} effects (\textit{right}). 
    The red line is the \ac{MCMC} Gaussian fit, while the vertical dashed blue lines indicate the rest frame transition wavelengths of the sodium doublet. The green background marks the exposures taken out-of-transit. The results are referred to the average spectrum after excluding HD1-N6, HD1-N8 and HD1-N9.}
    \label{fig:HD189733}
\end{figure*}

Five targets in the sample -- HD 189733 b, KELT-9 b, KELT-20 b, WASP-69 b, and WASP-76 b -- present a significant detection on most of the nights here analysed. For these targets, 
we report the results obtained from the \ac{MCMC} fit, for each night and when combining all nights (Figs. \ref{fig:HD189733}, \ref{fig:Kelt9}-\ref{fig:WASP-76}, \href{https://zenodo.org/records/14268745}{Tables B.2-B6}). Table \ref{tab:absorption_depths} shows 
the \acp{AD} calculated on three different central passbands (see Sect. \ref{sec:mcmc_analysis}). 
We point out that we expect the smallest passband (2 $\times$ 0.75 \AA) to produce the strongest detection, since most of the signal we measure originates in the narrow line cores ($\leq1 \, \AA$). Because some of the nights analysed had no signals in absorption in either of the sodium D lines or in only one line, we decided to repeat the analysis excluding these nights. Both pre-selection and post-selection results are presented below.

The reported results were obtained by setting a Gaussian prior on the $K\rm_p$ value approximately equal to the theoretical one, with a standard deviation of 10 km$\,$s$^{-1}$. 
However, without the prior, we verified that 
the contrasts and the FWHM values found on nights with detection are compatible with the ones found when the prior is set. This is not true only for 
some nights, like
HD1-N1, in which a deeper signal ($\sim 1\%$) is found at a $K\rm_p$ compatible with zero, and thus attributable to a stellar residual. Therefore, trying to exclude possible signals of non-planetary origin, we decided to evaluate the results found employing the prior on $K\rm_p$.

\begin{itemize}
    \item HD 189733 b. For this target, we measured the final contrasts of -0.32 $\pm$ 0.04 \%
    for the D$_2$ line and -0.29 $\pm$ 0.04 \%
    for the D$_1$ line, and a FWHM of 
    34.9$^{+4.5}_{-3.9}$ km$\,$s$^{-1}$. 
    Analysing the individual nights, we found that 
    HD1-N8 and HD1-N9
    report a very large uncertainty for both D lines (see Fig. \ref{fig:HD189733}, first and second upper panels), and a FWHM compatible with zero (see \href{https://zenodo.org/records/14268745}{Table B.2}). 
    This is
    probably attributable to the high 
    relative humidity for HD1-N8, and the presence of calima and variable seeing for HD1-N9. Besides, HD1-N6 presents an almost flat transmission spectrum, having very low contrasts ($\sim$ 0.1 \%) and a FWHM of $\sim$ 80 km$\,$s$^{-1}$. It should be noted that for this night it was not possible to check, and eventually remove, the presence of telluric sodium emission often found in La Palma, where HARPS-N is located, which may therefore have contaminated the transmission spectrum.
    Excluding these three nights from the analysis,
    we found final contrasts of -0.42 $\pm$ 0.05 \% for the D$_2$ line and -0.37 $\pm$ 0.05 \% for the D$_1$ line, and a FWHM of 33.3$^{+3.4}_{-3.0}$ km$\,$s$^{-1}$.
    Both $\tilde{\mathfrak{R}}$
    and the TLC after the selection are shown in Fig. \ref{fig:HD189733} (bottom panels). 
    The binned transmission spectrum shows a peculiar w-shaped profile in correspondence with both the \ion{Na}{I} absorption lines. This distortion is likely due to an imperfect correction for the main stellar contributions, which are the \ac{CLV} and the \ac{RM} effects \citep{Borsa-Zannoni}. 
    Specifically for this target, the main spurious contribution is due to
    stellar rotation, whose effect is condensed very closely to the centre of the spectral lines, where the planetary absorption mainly takes place \citep{keles2024pepsi}. 
    On the other hand, the impact of the \ac{CLV} effect, weaker in magnitude but persistent in a much greater wavelength range, and thus more evident on the TLC, seems to be more accurately modelled.
    In conclusion, we cannot exclude that the depth of the lines measured by our \ac{MCMC} fit may be underestimated due to these residuals, affecting the direct comparability with other planets with detections.
    
    Concerning the wind velocity extracted from the \ac{MCMC} fitting procedure (Fig. \ref{fig:HD189733}, third upper panel), we found a net blueshift with respect to the expected wavelength position of the \ion{Na}{i} lines of $\sim$ 6 km$\,$s$^{-1}$ in the case in which we combine all nights and 
    $\sim$ 5 km$\,$s$^{-1}$ in the case in which we exclude HD1-N6, HD1-N8, and HD1-N9.\\ 
    Looking at the \acp{AD} (Table \ref{tab:absorption_depths}), the strongest detection is 
    at 14-15$\sigma$ from $\tilde{\mathfrak{R}}$,
    and 5-6$\sigma$ from the TLC. The values found with the two approaches are not compatible and, in particular, the \acp{AD} extracted from $\tilde{\mathfrak{R}}$ are always higher and more precise. This is expected, since in one case ($\tilde{\mathfrak{R}}$) we are measuring the depth of the line, and in another (TLC) the missing flux.
    In any case, after excluding the three nights discussed above, 
    we found deeper absorption signals. \\
    The atmospheric \ion{Na}{I} absorption observed during the HARPS transits (first three nights of our dataset) has been revisited by several studies (e.g. \citealt{Wyttenbach_2015, CB_2017, Langeveld_2021}). 
    The use of different techniques for data reduction and analysis leads to results that are not always compatible. In this work, we have added eight more nights to those analysed in the literature (see Sect. \ref{sec:hd189}). Even if three of these nights  lead to a null detection, we can confirm the presence of \ion{Na}{I} in the planet's atmosphere
    with more accurate results, as the error on the average spectrum decreases with the number of observations. 
\\

    \item KELT-9 b. For this target, we found 
    the same contrast for both D lines
    in $\tilde{\mathfrak{R}}$: -0.13 $\pm$ 0.02 \%.
    When we combine all nights, the FWHM is 
    18.3$^{+2.2}_{-2.0}$ km$\,$s$^{-1}$, 
    while the shift from the rest frame wavelength of the sodium doublet is $\sim$ 
    -7 km$\,$s$^{-1}$.
    The absorption signal is present on all six nights analysed 
    (see \href{https://zenodo.org/records/14268745}{Table B.3} and Fig. \ref{fig:Kelt9}). In K9-N1, K9-N4, and K9-N6, one of the D lines contrasts is lower than 0.10 \%. In particular, during K9-N6 (the night with the lowest \ac{SN}), the D1 line is almost flat. Despite this, we decided not to exclude these nights from the analysis, as the uncertainties on the measurements are such that they do not cancel the signal.\\
    Considering the narrowest bands
    , we extracted a final \ac{AD} at $\sim$ 11$\sigma$ from $\tilde{\mathfrak{R}}$, and at $\sim$ 4$\sigma$ from the TLC (see Table \ref{tab:absorption_depths}).\\
    From the bottom left panel of Fig. \ref{fig:Kelt9}, a pattern is evident in the continuum of $\tilde{\mathfrak{R}}$,
    close to the D$_2$ line. We verified that it comes from K9-N1; in fact, it is also present in the final transmission spectrum reported by L2022. This pattern (whose origin is not clear) could also explain the two binned out-of-transit points in the TLC that present an offset with respect to the continuum  (see bottom right panel of Fig. \ref{fig:Kelt9}). 
    Neither the two points in the TLC nor the pattern in $\tilde{\mathfrak{R}}$ are present if we exclude K9-N1.\\
    Our results are in agreement with \citet{D'Arpa_k9}, who analysed our same dataset, and L2022, who analysed only K9-N1 and K9-N3. With the same observations, using the cross-correlation technique, \citet{Hoeijmakers_2019} detected (> 5$\sigma$) absorption by \ion{Na}{I}, later confirmed by \citet{Borsato_2023, Borsato_2024}.      
    For this target, the presence of \ion{Na}{I} D absorption features in the transmission spectrum is also predicted by \citet{Fossati_2021}. However, the measured absorption strength appears to be shallower than that predicted by models computed both assuming local thermodynamical equilibrium (LTE) and by accounting for non-local thermodynamical equilibrium (NLTE) effects, though the latter have been shown to reproduce well the observed hydrogen Balmer lines and the \ion{O}{I} nIR triplet \citep{Fossati_2021, Borsa_2022}. The difference in absorption strength between the observed and synthetic \ion{Na}{I} D lines could be due to a sub-solar Na abundance in the planetary atmosphere or to a higher temperature in the lower and middle atmosphere, which would thus increase Na ionisation. However, we disfavour the second possibility because it would likely lead to worsening the fit of the \ion{H}{I} Balmer lines and of the \ion{O}{I} nIR triplet.\\
    

\begin{figure*}[ht]
 \centering
  \includegraphics[width=\textwidth]{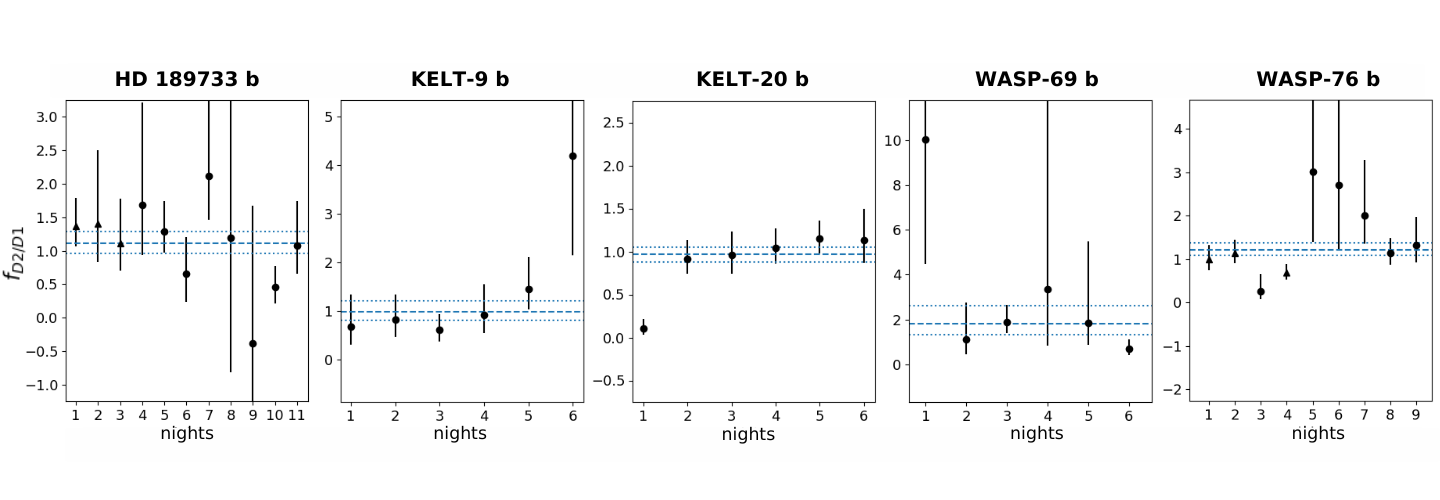}
    \caption{Variability of the line depth ratio for each target with detection. The average value and its 1$\sigma$ errors obtained by combining all nights are indicated by the horizontal dashed and dotted lines, respectively.}
    \label{fig:Dratio}    
\end{figure*}

 \item KELT-20 b. 
    In this case, too, the \ac{MCMC} analysis of $\tilde{\mathfrak{R}}$ converges on almost the same contrast value for the two sodium D lines; that is, -0.39 $\pm$ 0.03 \% for the D$_2$ line and -0.40 $\pm$ 0.03 \% for the D$_1$ line.
    Looking at the summary of the best-fit parameters (\href{https://zenodo.org/records/14268745}{Table B.4}) and Fig. \ref{fig:Kelt20}, we can see that all the nights analysed are characterised by an absorption signal in both D lines, only during K20-N1, the D$_2$ line shows a very low contrast ($\sim$ -0.06 \%) compared to other nights; this is likely due to the lower \ac{SN} (see \href{https://zenodo.org/records/14268745}{Fig. B.1}). Nevertheless, 
    we decided to keep this night in the ensemble analysis, for the same criterion applied to KELT-9 b nights.
    The FWHM is 14.6$^{+1.2}_{-1.0}$ km$\,$s$^{-1}$, while the shift of the Na D lines found by the \ac{MCMC} analysis is compatible with a net blueshift of $\sim$ 4 km$\,$s$^{-1}$.\\
    Concerning the \acp{AD} (Table \ref{tab:absorption_depths}), the most significant detection is at $\sim$ 22$\sigma$ from $\tilde{\mathfrak{R}}$ and $\sim$ 6$\sigma$ from the TLC, where the CLV+RM model closely matches the data (see Fig. \ref{fig:Kelt20}, bottom right panel). \\
    If we consider only the results relative to the combined nights, ours are compatible with \citet{CB_2019} and L2022, who analysed the first three nights of our dataset. Using the same data, \citet{Nugroho2020} confirmed the \ion{Na}{I} detection using the cross-correlation technique. 
    We compared the observed line profiles with the synthetic ones published by \citet{Fossati_2023}. We found that the synthetic transmission spectrum computed accounting for NLTE effects matches the observations well, while the synthetic transmission spectrum computed assuming LTE significantly underpredicts the strength of the planetary absorption lines.\\

     \item WASP-69 b. Compared to other targets with a sodium detection, this one shows a transmission spectrum characterised by a D$_2$ line much deeper than the D$_1$ line, as can be seen in Fig. \ref{fig:WASP69} (bottom left panel). The final contrasts in $\tilde{\mathfrak{R}}$ are -0.94$^{+0.20}_{-0.23}$ \% for the D$_2$ line and -0.51$^{+0.15}_{-0.15}$ \% for the D$_1$ line, while the FWHM is $\sim$ 20 km$\,$s$^{-1}$.
    Analysing the single nights (see \href{https://zenodo.org/records/14268745}{Table B.5} and upper panels of Fig. \ref{fig:WASP69}), during W69-N1 
    the D$_2$ line is detected with higher contrast than on other nights, but this value ($\sim$ -6 \%) is in agreement with what is reported by \citet{CB_2017} who analysed the same night, along with W69-N2.
    From our results, we found that W69-N4 and W69-N5 are characterised by a very narrow FWHM ($\sim$ 1-2 km$\,$s$^{-1}$). Besides, W69-N5 presents an excessive redshift; 
    this is not expected from global circulation models of the atmospheres of HJs, which predict the presence of zonal winds flowing from the planetary day-side to the night-side (e.g. \citealt{Komacek-Showman, Parmentier-Crossfield-book, Roman2021}).
    Excluding these two nights, we got deeper contrasts (-1.45$^{+0.25}_{-0.25}$ \% for the D$_2$ line and -0.89$^{+0.20}_{-0.22}$ \% for the D$_1$ line) and a higher blueshift with respect to the expected wavelength position of the \ion{Na}{i} lines ($v\rm_{wind} \sim$ -5 km$\,$s$^{-1}$).\\
    The most significant detection for the \acp{AD} is at $\sim$ 9-10$\sigma$ from $\tilde{\mathfrak{R}}$ and $\sim$ 3$\sigma$ from the TLC, where the \ac{CLV} effect is not clearly distinguishable. This time, the \ac{AD} values found with the two approaches are compatible within the uncertainties.\\
    In addition to \citet{CB_2017}, who reported an excess absorption only at the Na D$_2$ line in the passband of 1.5 \AA, at the level of 5$\sigma$, there have been other attempts to characterise the atmosphere of WASP-69 b. 
    Neither \citet{Deibert_2019} with the Gemini Remote Access to CFHT ESPaDOnS Spectrograph (GRACES)/Gemini nor \citet{Murgas_2020} with the Optical System for Imaging and low-Intermediate-Resolution Integrated Spectroscopy (OSIRIS)/ Gran Telescopio Canarias (GTC) found any signs of excess absorption at the sodium doublet wavelengths.  
    On the contrary, sodium detection in both D lines was obtained by \citet{Khalafinejad_2021}, who performed a combined high- and low-resolution transmission spectroscopy analysis, and L2022, who analysed only the first two nights of our datasets.\\

\begin{table*}[ht]
\caption{\label{tab:absorption_depths}Measured relative absorption depth in [\%] of atmospheric sodium on the targets with detection, extracted from the combined transmission spectrum ($\tilde{\mathfrak{R}}$) and from the TLC. Values are meant to be negative.}
 \centering
    \begin{tabular}{c| c c c| c c c}
    \hline
    \hline
     & \multicolumn{3}{c|}{$\tilde{\mathfrak{R}}$} & \multicolumn{3}{c}{TLC}\\
     & 0.75 \AA & 1.50 \AA & 3.00 \AA & 0.75 \AA & 1.50 \AA & 3.00 \AA \\
     \hline
    & \multicolumn{6}{c}{All nights}\\
    \hline
    HD 189733 b & 0.244 $\pm$ 0.017 & 0.127 $\pm$ 0.010 & 0.069 $\pm$ 0.006 & 0.149 $\pm$ 0.027 & 0.066 $\pm$ 0.019 & 0.025 $\pm$ 0.012\\
    KELT-9 b & 0.078 $\pm$ 0.007 & 0.034 $\pm$ 0.005 & 0.017 $\pm$ 0.004 & 
    0.049 $\pm$ 0.013 & 0.026 $\pm$ 0.009 & 0.021 $\pm$ 0.007\\
    KELT-20 b & 0.195 $\pm$ 0.009 & 0.116 $\pm$ 0.007 & 0.066 $\pm$ 0.005 & 0.134 $\pm$ 0.023 & 0.088 $\pm$ 0.017 & 0.055 $\pm$ 0.013\\
    WASP-69 b & 0.490 $\pm$ 0.054 & 0.261 $\pm$ 0.032 & 0.142 $\pm$ 0.020 & 0.383 $\pm$ 0.121 & 0.223 $\pm$ 0.073 & 0.144 $\pm$ 0.044\\ 
    WASP-76 b & 0.307 $\pm$ 0.018 & 0.164 $\pm$ 0.013 & 0.070 $\pm$ 0.009 & 0.224 $\pm$ 0.041 & 0.147 $\pm$ 0.029 & 0.069 $\pm$ 0.022\\
       \hline
    & \multicolumn{6}{c}{After selection}\\
    \hline
    HD 189733 b &  0.300 $\pm$ 0.020 & 0.155 $\pm$ 0.012 & 0.073 $\pm$ 0.007 & 0.206 $\pm$ 0.032 & 0.099 $\pm$ 0.022 & 0.042 $\pm$ 0.014\\
    KELT-9 b & - & - & - & - & - & -\\
    KELT-20 b & - & - & - & - & - & -\\
    WASP-69 b & 0.749 $\pm$ 0.074 & 0.373 $\pm$ 0.044 & 0.218 $\pm$ 0.027 & 0.519 $\pm$ 0.168 & 0.324 $\pm$ 0.102 & 0.227 $\pm$ 0.061\\
    WASP-76 b & 0.306 $\pm$ 0.018 & 0.162 $\pm$ 0.013 & 0.069 $\pm$ 0.009 & 0.283 $\pm$ 0.037 & 0.178 $\pm$ 0.027 & 0.094 $\pm$ 0.020 \\
    \hline
    \hline
    \end{tabular}
\end{table*}

 \item WASP-76 b. Combining all nights, we got final contrasts of -0.52 $\pm$ 0.05 \% for the D$_2$ line and -0.43 $\pm$ 0.04 \% for the D$_1$ line, 
 and a FWHM of 22.3$^{+2.0}_{-1.9}$ km$\,$s$^{-1}$. As can be seen from the values listed in \href{https://zenodo.org/records/14268745}{Table B.6} and reported in Fig. \ref{fig:WASP-76}, the \ac{MCMC} fitting procedure did not converge in W76-N6 and the FMWH is compatible with zero; we point out that this night is the one with the lowest \ac{SN}. Besides, the contrasts measured during W76-N3 present larger uncertainties; this can be due to the high seeing during the night ($\sim$ 3"). Both nights are characterised by a higher $v\rm_{wind}$ than the value obtained by combining all nights ($\sim$ -4.5 km$\,$s$^{-1}$). In contrast, we found values compatible for all other nights analysed.
In the end, we decided to exclude only W76-N6 from the combined analysis, as the contrasts and FWHM found on W76-N3 are not compatible with zero. Applying the \ac{MCMC} analysis again, 
we did not derive any significant difference in the fitting parameters.
The same is true also for the \acp{AD} extracted from $\tilde{\mathfrak{R}}$, where 
 we found a detection of $\sim$ 17$\sigma$ at 0.75 \AA. Instead, the \acp{AD} measured from the TLC ($\sim$ 5.5$\sigma$) become more significant after excluding W76-N6 ($\sim$ 7.7$\sigma$).

 The first detection of \ion{Na}{I} in this target was reported by \citet{Seidel_2019}, who analysed the HARPS nights of our dataset. The same nights were analysed by L2022, along with other two transits retrieved with HARPS, not analysed in this work. Our results are in agreement with what they found.
 At high resolution, the detection was also confirmed by \citet{Tabernero_2021} using two transits with ESPRESSO, while only an upper limit was reported by \citet{CB-w76} with CARMENES.

\end{itemize}

\subsubsection{Line depth ratio}
For the targets of our sample characterised by a planetary absorption signal (see Sect. \ref{sec:detections}), we computed the line depth ratio, $f\rm_{D2/D1}$, 
for each night and for $\tilde{\mathfrak{R}}$.
This quantity is able to provide information on the average velocity distributions of the Na atoms, their ongoing behaviour, and which atmospheric regime the majority of the absorption occurs in \citep{Oza_2019, Gebek-Oza}.
We calculated it using the median values of the posterior distributions of the contrasts ($c$ D$_2$, $c$ D$_1$). The confidence interval between the 16th and 84th percentiles of the ratio of the distributions of the two single D-lines was used as error bars.
The results are reported in \href{https://zenodo.org/records/14268745}{Tables B.2 - B6}, while a summary of $f\rm_{D2/D1}$ obtained when combining all nights, and after excluding the nights without detection, is listed in Table \ref{tab:Dratio}.

Except for WASP-69 b, all the targets with sodium detection have an average ratio that is consistent with or slightly larger than one (within the uncertainties). 
 This suggests that these planets could share common atmospheric hydrostatic scenarios. 
This agrees with the predictions of forward modelling computed accounting for NLTE effects \citep{Fossati_2021, Fossati_2023}. 
On the contrary, WASP-69 b is characterised by an average line ratio greater than 1.5, which could be an indication of an evaporative, non-hydrostatic scenario \citep{Gebek-Oza}. 

Figure \ref{fig:Dratio} shows the variability of the line depth ratio for each target with detection. 
We can see that for KELT-9 b and KELT-20 b, the ratio seems to increase with time. 
This behaviour could be a sign that perhaps the atmospheric regime is undergoing some kind of change. However, the high uncertainties about the estimate of $f\rm_{D2/D1}$ on some of the nights analysed (e.g. K9-N6) do not allow us to come to a reliable conclusion about its atmospheric dynamics. 
KELT-9 b and KELT-20 b are also the targets with detection where we find the lowest average $f\rm_{D2/D1} \, (\leq$ 1). This is due to the fact that the first nights analysed are characterised by a deeper contrast in line D$_1$ than in line D$_2$.
For HD 189733 b, WASP-69 b, and WASP-76 b, there is variability in the line depth ratio but not a clear trend; anyway, most of the time, $f\rm_{D2/D1}$ is closer to one inside the uncertainties, for HD 189733 b and WASP-76 b, while closer to or compatible with two, for WASP-69 b.

Our results are in agreement with the line depth ratios reported by L2022 for the same targets. This was expected, since we used some of the same data, and similar methods for the analysis. For HD 189733 b, \citet{Gebek-Oza} estimate $f\rm_{D2/D1} = 1.74 \pm 0.45$, which is higher than our findings (even if compatible within the uncertainties). It is important to note that, for the measurement, they used the transit depths averaged over bandwidths of 0.2 \AA, centred on the absorption lines, combining the first three nights of our dataset. We verified that using only these nights, and computing the line depth ratio as they did, we find the same $f\rm_{D2/D1}$ value ($\sim 1.7$).

\begin{table}[ht]
\centering
\begin{spacing}{1.4}
 \caption{\label{tab:Dratio} Depth line ratio.}
    \begin{tabular}{c|c|c}
    \hline
    \hline
    Planet & $f_{D2/D1}$ & $f_{D2/D1}$$^{(*)}$ \\
    \hline
    HD 189733 b & 1.1$_{-0.2}^{+0.2}$ & 1.1$^{+0.2}_{-0.1}$  \\
    KELT-9 b & 0.99$_{-0.18}^{+0.21}$ & - \\
    KELT-20 b & 0.98$^{+0.08}_{-0.09}$ & - \\
    WASP-69 b & 1.8$_{-0.5}^{+0.8}$ & 1.6$_{-0.4}^{+0.5}$ \\
    WASP-76 b & 1.2$_{-0.1}^{+0.1}$ & 1.2$_{-0.1}^{+0.2}$ \\
    \hline
    \hline
    \end{tabular}
    \end{spacing}{}
\tablefoot{
        \tablefoottext{*}{After selection.}
}
\end{table}

\begin{figure*}
    \centering
    \includegraphics[width=\textwidth]{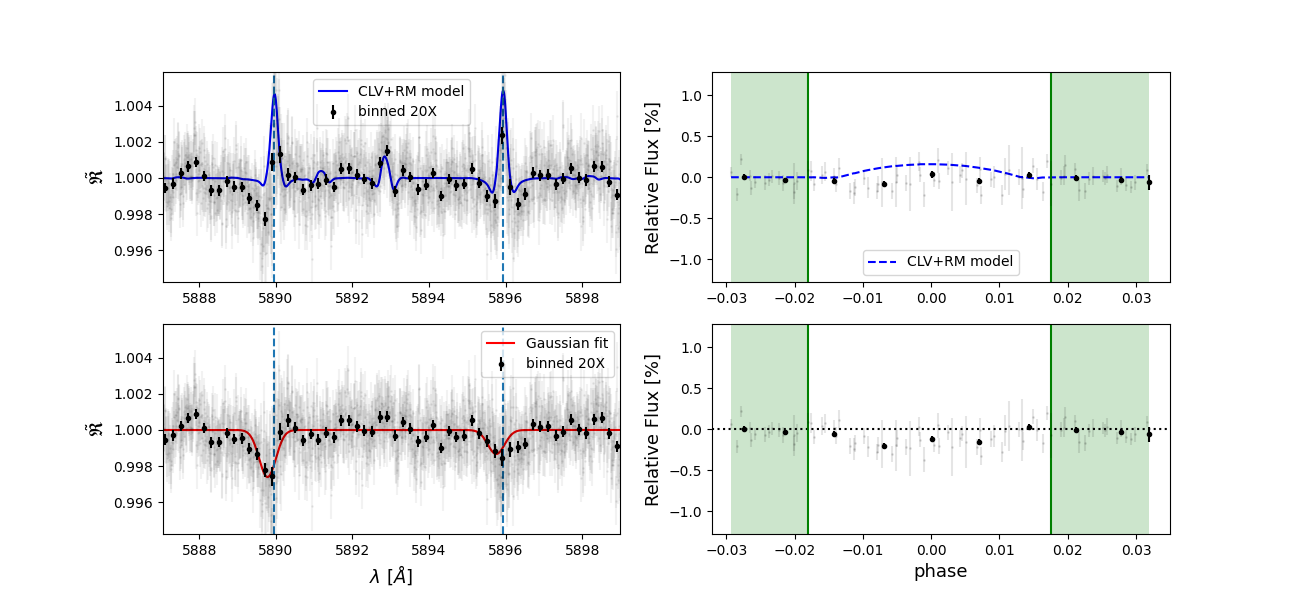}
    \caption{Final transmission spectrum (\textit{left}) and relative TLC (\textit{right}) of HD 209458 b without the correction of the \ac{CLV} and \ac{RM} effects (\textit{upper panels}) and after their correction (\textit{bottom panel}). The blue line is the best-fit CLV+RM model; the red line is the \ac{MCMC} Gaussian fit. The plots refer to the results obtained after excluding HD2-N1, HD2-N2, HD2-N3, HD2-N4, and HD2-N6.}
    \label{fig:HD209458}
\end{figure*}

\subsection{Targets with unclear detections}
\label{sec:residuals}
The last three targets of our sample, namely HD 209458 b, WASP-80 b, and WASP-127 b, deserve a separate discussion as the results of our analysis raise doubts on the planetary origin of the signals observed in their transmission spectrum. 

\begin{itemize}
    \item HD 209458 b. As we can see in the upper panel of Fig. \ref{fig:HD209458} (on the left) and as is reported by \citeauthor{CB_2020} (\citeyear{CB_2020}, \citeyear{CB_2021}), the transmission spectrum of HD 209458 b does not present an atmospheric absorption from the planet, but a pseudo-emission signal centred on the \ion{Na}{I} lines positions. This signal, visible on each individual night, is quite consistent with the model containing the \ac{CLV} and \ac{RM} contributions (represented with a blue line), not only for the sodium lines but also for other stellar lines (e.g. at around 5893 \AA, consistent with \ion{Ni}{I}).
    After subtracting the CLV+RM model from the transmission spectrum, some absorption features arise. Applying the \ac{MCMC} fitting analysis, we got final contrasts of 
    -0.16 $\pm$ 0.04 \% and -0.15 $\pm$ 0.03 \% 
    for the D$_2$ and D$_1$ lines, respectively, with a FWHM of 30.1$^{+7.1}_{-5.8}$ km$\,$s$^{-1}$ and a blueshift of 1.5$^{+1.8}_{-1.6}$ km$\,$s$^{-1}$.  
    However, 
    according to \citet{CB_2020}, the absorption-like signals visible in $\tilde{\mathfrak{R}}$ probably result from the combination of the limited model accuracy and the smaller \ac{SN} in the line cores. Indeed, this pseudo-absorption signal is not present (or it is compatible with zero) on most of the eight nights analysed (see \href{https://zenodo.org/records/14268745}{Table B.7}), except for HD2-N5, HD2-N7, and HD2-N8.
    Combining them the absorption signal becomes more significant in D$_2$ ($\sim 6\sigma$), and less significant in D$_1$ ($\sim 3\sigma$).
  
    In terms of the \ac{AD},
    while the measurement extracted from the TLC is significant only at 3$\sigma$ in the widest passband, the one extracted by $\tilde{\mathfrak{R}}$ ($\sim$ 10 $\sigma$) suggests a planetary absorption. 
    
    Our results are in agreement with what was measured by \citet{CB_2020}, who analysed the first three nights of our dataset, combined with CARMENES observations. 
   With ESPRESSO, \citet{CB_2021} confirmed the hypothesis that the \ion{Na}{I} absorption features observed in the transmission spectrum are not from the exoplanet atmosphere. Rather, the most likely origin is the deformation of the stellar line profile due to the transit, mainly produced by the \ac{RM} effect. 
   To further support this, they found
   that signals are present at all the wavelengths where the host star shows strong absorption lines (e.g. \ion{Fe}{I}, \ion{Ca}{I}, H$\alpha$), 
    using both single-line analysis and the cross-correlation technique. A similar result was obtained in another gas giant, the planet HAT-P-67 b  \citep{Sicilia_hatp67}.\\

    
 

\item WASP-80 b. 
Of the seven nights 
of our dataset (see \href{https://zenodo.org/records/14268745}{Fig. B.2} and \href{https://zenodo.org/records/14268745}{Table B.8} for the combined analysis), three nights (W80-N1, W80-N5, and W80-N7) are characterised by contrasts or FWHM compatible with zero in their transmission spectra, while on two nights (W80-N2 and W80-N4) the \ac{MCMC} analysis fits a signal in emission in both D lines. 
The only nights of our dataset on which we were able to fit an absorption signal are W80-N3 and W80-N6. However, we verified that this signal is also present in the average-out spectrum of W80-N3, which is the transmission spectrum built using only out-of-transit spectra. Indeed, by comparing all the master out spectra of each night (see \href{https://zenodo.org/records/14268745}{Fig. B.3}), an excess of emission is evident on this night.
Considering only W80-N6, we found contrasts of $\sim$ -4 \% and -7 \% for the D$_2$ and D$_1$ lines, respectively (see Fig. \ref{fig:WASP-80}). The very deep signals found and the unusual line depth ratio (D$_1$ line deeper than the D$_2$ line), along with the consideration that this is the only night out of seven on which there seems to be a detection, lead us to rule out a planetary origin. 
It is much more plausible to think that the sodium absorption in W80-N6 is due to a variable level of chromospheric activity in the host star. Actually, according to \citet{Fossati_w80}, WASP-80 has a colour index of B-V = 1.34, which is larger than the 1.1 limit beyond which the \ion{Na}{I} D lines become sensitive to the level of chromospheric activity according to \citet{Diaz_2007}.\\
It should also be noted that this target is one of the faintest (V $\sim$ 12) to be studied with transmission spectroscopy (at least with a 3.5 m-class telescope). Along with the low number of in-transit spectra (see \href{https://zenodo.org/records/14268745}{Table B.1}), this makes the detection of a possible planetary signal more difficult. As was done for GJ 436 b (Sect. \ref{sec:no-detections}), we verified that the smearing effect deriving from the long exposure ($\sim$ 3.4 pixels) can be neglected, while the Doppler shift due to the variation in the planetary velocity during the transit (from $\sim$ -10 to +10 km$\, $s$^{-1}$, corresponding to $\sim$ 0.20 \AA) would keep the signal located in the low \ac{SN} region of the stellar line cores.\\
Our results are in agreement with \citet{Parviainen_2018}, who recovered a flat transmission spectrum with no evidence of \ion{Na}{I} absorption using the OSIRIS spectrograph. For this target, there is also an earlier tentative detection by \citet{Sedaghati_2017} at low resolution with FORS2 (FOcal Reducer/low dispersion Spectrograph 2).
\\

\item WASP-127 b.
Combining all nights available for this target, we got final contrasts of -0.18 $\pm$ 0.03\% and -0.17 $\pm$ 0.03\%
for the D$_2$ and D$_1$ lines, respectively (>5$\sigma$). However, the values found for the FWHM and the $v_{wind}$ value are too high ($\sim$ 91 km$\,$s$^{-1}$ and $\sim$ 18 km$\,$s$^{-1}$, respectively) to ascribe this absorption to the planetary atmosphere. Besides, 
$\tilde{\mathfrak{R}}$ turns out to be redshifted with respect to the nominal rest wavelengths of the \ion{Na}{i} lines. \\
In all the six nights analysed for this target, the \ac{MCMC} analysis is able to fit a double absorption signal (see \href{https://zenodo.org/records/14268745}{Table B.9}), even if not significant enough (<3$\sigma$). In particular, in W127-N4 the signal detected is at the noise level 
present in the region of the sodium doublet, and W127-N5 is characterised by an emission feature in correspondence of the D$_1$ line. We remark that W127-N4 and W127-N5 were affected by moonlight and a very variable seeing, respectively. Excluding both nights, along with W127-N3, which presents an almost flat and highly redshifted transmission spectrum, we got a less significant detection (<5$\sigma$), but with an FWHM and $v_{wind}$ value consistent with those of a possible planetary signal (see Fig. \ref{fig:WASP127}).
We must emphasise that our selection criterion, which led to the exclusion of half of the nights available for this target, may have affected the detection, which remains unclear. Further investigation with more observations, preferably at a higher \ac{SN}, would be needed to clarify the origin of the signal.

With the first two nights of our dataset, \citet{Zak} reported the detection of sodium in the atmosphere of WASP-127 b at a 4–8$\sigma$ level of significance, confirming earlier results based on low-resolution spectroscopy \citep{Palle_2017,Chen_2018}. Later, \citet{Seidel_127} showed that this detection was actually due to contamination from telluric sodium emissions and the low \ac{SN} in the core of the deep stellar sodium lines. Their findings were confirmed by \citet{Sicilia}. 
However, both at high resolution with ESPRESSO \citep{Allart_2020}, and at low resolution combining data HST/STIS and HST/WFC3 \citep{Spake_2020,Skaf_2020}, the sodium line core was detected at 9$\sigma$ and 5$\sigma$, respectively.

\end{itemize}

\begin{table}
\centering
\begin{spacing}{1.4}
 \caption{\label{tab:relative_heights} Relative heights found in this work,  
 compared with the ones obtained by L2022 \citep{Langeveld_2022}.}
    \begin{tabular}{c ccc}
    \hline
    \hline
    & \multicolumn{3}{c}{$H_{D2}/R_p$}\\
    Planet & This work &This work $^{(*)}$ & L2022\\
    \hline
    HD 189733 b & 0.069$_{-0.010}^{+0.010}$ & 0.090$_{-0.012}^{+0.012}$ & 0.078$\pm$0.013\\
    KELT-9 b & 0.092$_{-0.014}^{+0.014}$ & - & 0.107$\pm$0.019\\
    KELT-20 b & 0.131$_{-0.016}^{+0.016}$ & -&0.107$\pm$0.017\\
    WASP-69 b & 0.235$_{-0.057}^{+0.052}$ & 0.346$_{-0.062}^{+0.061}$&0.684$\pm$0.124\\
    WASP-76 b & 0.201$_{-0.027}^{+0.027}$ & 0.199$_{-0.027}^{+0.027}$&0.178$\pm$0.023\\
    \hline
    HD 209458 b & 0.053$_{-0.013}^{+0.013}$ & 0.085$_{-0.013}^{+0.013}$&-\\   
    WASP-80 b & 0.090$_{-0.091}^{+0.082}$ & 0.489$_{-0.201}^{+0.184}$&-\\
    WASP-127 b & 0.086$_{-0.016}^{+0.015}$ & 0.227$_{-0.061}^{+0.052}$&-\\
    \hline
    \hline
    \end{tabular}
    \end{spacing}{}
\tablefoot{
        \tablefoottext{*}{After selection.}
}
\end{table}

\section{Interpretation of the transmission spectrum}\label{sec:interpretation}
The formulae that will be described below are introduced in more detail in L2022, in their Sect. 5.

The absorption depth measured at a specific wavelength ($\delta_\lambda$)
can be related to a correspondent height of the exoplanet's atmosphere ($H_\lambda$), according to the following relation:

\begin{equation}\label{eq:delta}
 \delta_\lambda = \frac{2\Delta_0H_\lambda}{R_p} + \Delta_0\biggl(\frac{H_\lambda}{R_p}\biggr)^2, 
\end{equation}
where $\Delta_0$ is the white-light transit depth, defined as $(R_p/R_\star)^2$. In our case, we are interested in converting the contrast of the sodium D lines measured from the \ac{MCMC} fit, 
especially for the targets with (unclear)detections (see Sects. \ref{sec:detections} and \ref{sec:residuals}).
In order to take into account the different $\Delta_0$ of each planet in the sample, and for a better comparison with L2022, we evaluated the relative height; that is, the ratio of $H_\lambda$ to $R\rm_p$. Considering Eq. \ref{eq:delta}, this is given by

\begin{equation}\label{eq:relative_height}
    \frac{H_\lambda}{R_p} \equiv h_\lambda = \sqrt{1+\frac{\delta_\lambda}{\Delta_0}}-1.
\end{equation}

\begin{figure}
    \centering
    \includegraphics[width=\columnwidth]{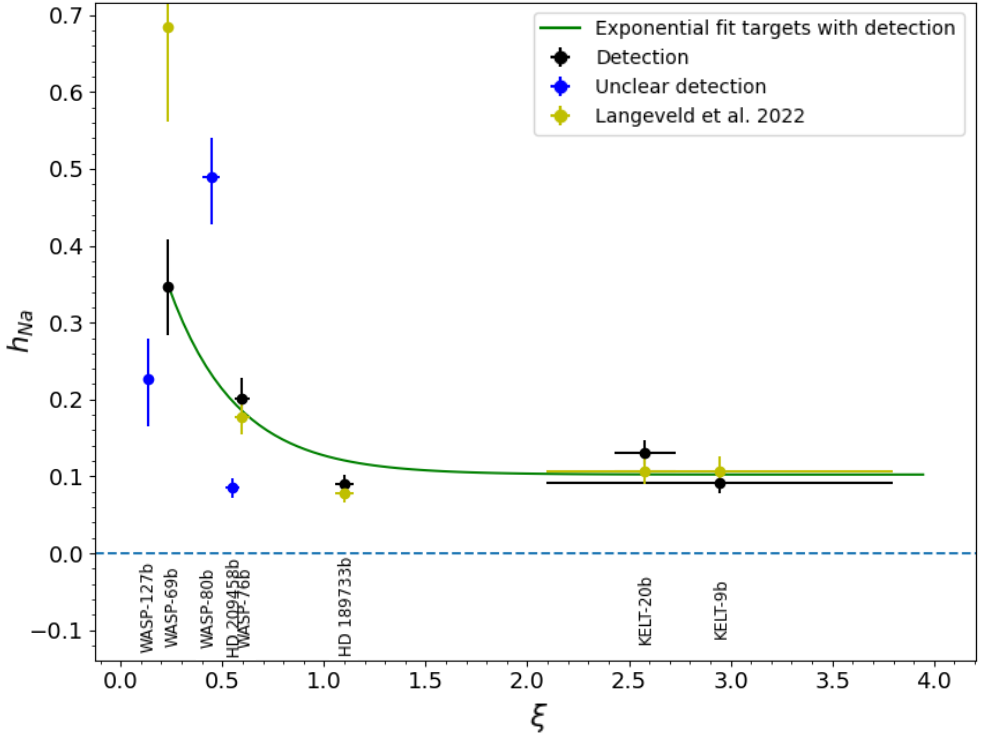}
    \caption{Relative atmospheric height of sodium as a function of the scaled quantity $\xi$. Black markers indicate the values obtained for the targets of our sample with detection, while blue markers are the ones obtained for targets with residuals in $\tilde{\mathfrak{R}}$. The exponential curve (green line) was fitted only to the targets with detection. For comparison, results reported by \citet{Langeveld_2022} for the same targets are also indicated (yellow).}
    \label{fig:relative_height}
\end{figure}

Assuming $\delta_\lambda$ equal to the measured $c$ D$_2$, we computed the relative atmospheric height of the sodium layer ($h_\lambda = h_{Na}$), before and after the selection.
The results are listed in Table \ref{tab:relative_heights}. In general, our results are consistent with the values found by L2022, except for WASP-69 b, for which they analysed only the first two nights of our dataset, the ones with the higher $c$ D$_2$ value.

L2022 proposed an additional method of assessing detectable Na features from high-resolution observations. They found that $h_{Na}$ follows an empirical exponential trend with the quantity:

\begin{equation}
    \xi = \biggl(\frac{T_{eq}}{1000 \, K}\biggr) \, \biggl(\frac{g}{g_J}\biggr), 
\end{equation}
where $T\rm_{eq}$ is the equilibrium temperature and $g/g\rm_J$ is the planet's surface gravity in Jovian units. This means that sodium detection should be more significant on planets with lower $\xi$ values. 
Using their $\xi$ values for the common planets (see their Table 4), we tried to verify the same empirical law:
\begin{equation}\label{eq:exp}
    h_{Na} =a \mathrm{e}^{-b \xi} + c. 
\end{equation}

Figure \ref{fig:relative_height} shows our findings when considering only the nights of our dataset with a significant detection. Anyway, we checked that similar results (but with higher uncertainties) are obtained when we consider all nights. By using the \texttt{curve\_fit} function from \texttt{scipy.optimize}, we fitted
the results relative to the targets with detection (see Sect. \ref{sec:detections}). We actually found the exponential trend indicated in Eq. \ref{eq:exp}, with the following best-fit values: $a=0.495 \pm 0.154$, $b=2.99 \pm 1.22$ and $c=0.102 \pm 0.022$, the last one being the horizontal asymptote of the exponential function. This value represents an upper limit on $h_\lambda$ for targets with $\xi$ values larger than $\sim$1.5. We also tried to include in the fit the targets that 
are affected by variability or final residuals
(see Sect. \ref{sec:residuals}), marked in blue in the figure, obtaining compatible best-fit values but with very high uncertainties. However, while the final relative height measured for WASP-127 b seems to follow the exponential trend, this is not true for HD 209458 b and WASP-80 b. Besides, we point out that if we consider all nights, WASP-80 b present a relative height compatible with zero; this suggests that the excess of absorption found for this planet after the selection is probably not of planetary origin. Likewise, according to the exponential law, we would have expected a higher $h_\lambda$ for HD 209458 b. Instead, we can see that, both by combining all nights and after selection, the value obtained is underestimated. Given the high quality of the dataset used for this last target, the lower detection could be a symptom of some physical processes in the planet's atmosphere. Recently, \citet{Canocchi2024_new} found that using one-dimensional NLTE stellar models for the correction of the \ac{CLV} and \ac{RM} effects on this target results in residuals that can be mistaken for a sodium absorption feature corresponding to a depth of about 0.2 \% (compatible with our findings, see \href{https://zenodo.org/records/14268745}{Table B.7}). On the other hand, three-dimensional NLTE stellar models can explain the spectral features found without the need for any planetary absorption. However, sodium could still arise from the planetary atmosphere
of HD 209458 b, more particularly from the troposphere, where the wings of the alkali doublets are generated and visible at median resolution \citep{Pino_code}, but removed at high altitudes by unknown processes, such as condensation or ionisation (e.g. \citealt{Vidal-Madjar2011}).

Using the fitted exponential trend,
we also calculated the expected line depth for the planets in our sample without detection (Sect. \ref{sec:no-detections}). The relative height found for GJ 436 b (at $\xi \sim 0.41$) and KELT-7 b (at $\xi \sim 1.2$) should correspond to a line contrast of $\sim$ 0.41 \% and 0.20 \%, respectively. 
Despite the fact that these values are compatible, or even larger, than those found on some of the planets with detection (i.e. HD 189733 b, KELT-9 b and KELT-20 b), both the estimated line contrasts are well within the continuum noise seen in Figs. \ref{fig:GJ436} and \ref{fig:Kelt7}. 
Therefore, a non-detection for these two planets is not surprising, considering also the possible causes already discussed in Sect. \ref{sec:no-detections}. We do not exclude that additional data with much higher \ac{SN} could reveal the presence of \ion{Na}{I} in the exoplanets' atmosphere.

As L2022 showed, we verified that also with our data, assuming ideal conditions where observations probe the same number of scale heights and the terminator is isothermal with a temperature, $T\rm_{eq}$, for all planets, the correlation does not improve.
In fact, by comparing the $h_\lambda$ values obtained (using Eq. \ref{eq:relative_height}) with the theoretically expected values, we find that they deviate. The only one of the five targets with detection that is compatible (within uncertainties) with the expected value is WASP-69 b.
For this same target, we point out that the value we found differs from that obtained by L2022. Probably, the strong signal detected on the first two nights (the only ones analysed by L2022) is somewhat mitigated by the other nights analysed in this work, which are characterised by a weaker signal. We have verified that if we combine only the first two nights, the measured relative height (0.849$_{-0.286}^{+0.202}$) is compatible with the one measured by L2022.

\section{Summary and conclusions}\label{sec:conclusions}
The neutral sodium doublet is one of the most investigated atmospheric signatures. Thanks to its large absorption cross-section, this atomic species has been detected in the upper atmosphere of several gas giants.
In this work, we present a homogeneous search for \ion{Na}{I} absorption in ten favourable targets for high-resolution transmission spectroscopy. 


After extracting the transmission spectrum and the TLC in the region of the sodium doublet for each target in our sample, we analysed the signals found on each night with an \ac{MCMC} fitting procedure, using the public code \texttt{SLOPpy} \citep{Sicilia}. The data reduction includes the removal of telluric lines and the modelling of the \ac{CLV} and \ac{RM} effects. In total, we worked on more than 2,000 spectra, 
covering 71 transit events, retrieved with HARPS and HARPS-N in the framework of GAPS or other scientific programmes.

Two of the ten targets investigated, namely GJ 436 b and KELT-7 b, did not show any significant absorption signal on any of the nights analysed (despite the fact that the observations span a very long period of time). This can be interpreted as the possible presence of high-altitude clouds or hazes in their atmosphere. An alternative interpretation is that the planets may not have enough sodium in their atmospheres to be detected, or that the sodium is present but quickly ionised above the clouds.
In the case of KELT-7 b, the potential planetary signal could also be hidden by a stronger \ac{RM} effect or stellar pulsations. Our results are in agreement with previous works.

As for the other eight targets in our sample, we have found some variability in the \ion{Na}{I} signal. 
Five targets -- namely HD 189733 b, KELT-9 b, KELT-20 b, WASP-69 b, and WASP-76 b -- show a significant detection on most of the nights analysed. All of them have been extensively studied in the literature, and all already have atmospheric sodium detections. Unlike the other works, though, the large availability of data for each target allows us to probe the signal variability. In fact, on some of the nights analysed, the \ac{MCMC} fit fails to converge on a significant enough absorption signal, even finding a flat transmission spectrum in some cases. After closer investigation, we realised that these nights very often are characterised by bad atmospheric conditions (e.g. variable seeing), lower \ac{SN}, or sub-optimal data treatment (e.g. sky spectrum not available). In some cases, the variability found seems to depend instead on the higher stellar activity of the host star.
Indeed, both for HD 189733 b and WASP-69 b, Guilluy et al. (\citeyear{Guilluy_2020, Guilluy2024}) report
a night-to-night variation in the Helium triplet absorption signal, associated with variability in H$\alpha$, which likely indicates the influence of pseudo-signals related to stellar activity.
Along with \ion{Ca}{II} H\&K, H$\alpha$ is one of the most useful stellar activity diagnostics in the \ac{VIS} range (e.g. \citealt{Gomes_da_Silva2022}). 
While simultaneous monitoring of this spectral line is beyond the scope of our paper, the correspondence of the results reported in the literature with the hints of variability found in our targets hosted by active stars seems to support the influence of stellar activity on the atmospheric signal.

For these five targets, we derived the line depth ratio of the sodium doublet $f\rm_{D2/D1}$, in order to understand in which atmospheric regime the majority of the absorption occurs. Despite the contrast variability in the D$_2$ and D$_1$ lines found on different nights, our results suggest a non-hydrostatic scenario only for WASP-69 b, while the other four planets in which we confirmed the detection seem to share a common hydrostatic atmosphere.

Finally, for three of the targets in our sample -- namely, HD 209458 b, WASP-80 b, and WASP-127 b -- we cannot confirm the presence of sodium in their atmospheres. Indeed, while most of the nights analysed are characterised by emission features or a featureless transmission spectrum, only on some nights do the final transmission spectra ($\tilde{\mathfrak{R}}$) and light curves present some absorption features.

Following \citet{Langeveld_2022}, who performed the first survey of sodium absorption (the five targets with detection in our sample are in common with theirs), we converted the absorption depth measured from $\tilde{\mathfrak{R}}$ to a correspondent relative atmospheric height. With respect to their work, we have a higher number of observations for the same targets analysed, and an additional target with a lower $\xi$ value (WASP-127 b, $\xi$=0.134), which seems to follow 
the same exponential trend as the other targets that present a significant detection. 
Given the values of the equilibrium temperature and surface gravity of HD 209458 b and WASP-80 b, the relative height we found is underestimated and overestimated, respectively. This leads us to ascribe the absorption signals found on their transmission spectra to probable residuals due to imperfect removal of the stellar contribution or to the smaller \ac{SN} in the line cores. 
On the contrary, the sodium absorption measured on WASP-127 b after the selection (-0.52$^{+0.13}_{-0.15}$ \% and -0.36$^{+0.10}_{-0.11}$ \% for the D$_2$ and D$_1$ lines, respectively) is likely to have a planetary origin.
Finally, we computed the  expected signal for the two targets without detection (i.e. GJ 436 b and KELT-7 b), assuming they follow the same empirical trend. 
The observed signals are well below the noise level, despite the expected values being large enough to be detectable with HARPS/HARPS-N.
In addition to the possible scenarios listed above for the non-detection, another explanation could be that the atmospheric signal on these planets is simply hidden by noise, whatever its origin (stellar contamination, instrumental effects, etc.). In conclusion, we cannot exclude that observations with a higher \ac{SN} would help to make an eventual atmospheric signal detectable on these two planets.

In addition to emphasising the importance of having many transit events for the same target, this work suggests that the apparent variability of atmospheric sodium (when present) does not appear to be related to intrinsic properties of the planet (at least in the cases studied here), but rather to observational conditions or limitations in data reduction techniques. However,
given the high uncertainties found in some measurements, we cannot completely rule out that the variability of the detected signal may in some cases be due to the planet's atmosphere itself. 

\section{Data availability}
Additional material including Tables B.1-B.9 and Figs. B.1-B.3 can be found on Zenodo: \href{https://zenodo.org/records/14268745}{https://zenodo.org/records/14268745}.

\begin{acknowledgements}
We thank the anonymous referee for their interesting and useful comments that helped improve the clarity of the paper. 
The authors acknowledge financial contribution from PRIN INAF 2019 and from the European Union - Next Generation EU RRF M4C2 1.1 PRIN MUR 2022 project 2022CERJ49 (ESPLORA), and the INAF GO Large Grant 2023 GAPS-2. They also acknowledge the Italian center for Astronomical Archives (IA2, https://www.ia2.inaf.it), part of the Italian National Institute for Astrophysics (INAF), for providing technical assistance, services and supporting activities of the GAPS collaboration. D.S. acknowledges the funding support from CHEOPS ASI-INAF Agreement n. 2019-29-HH.0 and Addendum n. 2019-29-HH.1-2022. D.S. thanks L. Mancini, V. Singh and T. Zingales for their comments and insights.    
\end{acknowledgements}

\bibliographystyle{aa}
\bibliography{sample}

\appendix
\onecolumn
\section{Additional figures and tables}

\begin{table*}[h]
 \caption{\label{tab:pams1}GJ 436, HD 189733, HD 209458, KELT-7, and KELT-9 system parameters adopted in this work.}
 \centering
 \begin{spacing}{1.4}
 \scriptsize
        \begin{tabular}{l c c c c c}
                \hline \hline
   & GJ 436 b & HD 189733 b & HD 209458 b & KELT-7 b & KELT-9 b\\
  \hline
  \underline{Stellar Parameters} \\
        $M_\star$ [M$_\sun$] &  0.445 $\pm$ 0.044 $^{(a)}$& 0.812$^{+0.041}_{-0.038}$ $^{(e)}$ & 1.119 $\pm$ 0.033 $^{(i)}$& 1.517 $\pm$ 0.022         $^{(n)}$& 2.32 $\pm$ 0.16 $^{(p)}$\\
        $R_\star$ [R$_\sun$] & 0.449 $\pm$ 0.019 $^{(a)}$ & 0.765$^{+0.019}_{-0.018}$ $^{(e)}$& 1.155$^{+0.014}_{-0.016}$ $^{(i)}$& 1.712 $\pm$ 0.037 $^{(n)}$ & 2.418 $\pm$ 0.058 $^{(p)}$\\
        $T\rm_{eff}$ [K] & 3354 $\pm$ 110 $^{(b)}$& 5050 $\pm$ 20 $^{(e)}$& 6118 $\pm$ 25 $^{(j)}$& 6699 $\pm$ 24 $^{(n)}$& 9600 $\pm$ 400 $^{(p)}$\\
        $[\rm Fe/H]$ [dex] &  -0.03 $\pm$ 0.09 $^{(b)}$& -0.08 $\pm$ 0.03 $^{(f)}$ & 0.03 $\pm$ 0.02 $^{(j)}$& 0.24 $\pm$ 0.02 $^{(n)}$ & 0.14 $\pm$ 0.30 $^{(p)}$\\
    $\log g$ [$\log_{10}$(cm s$^{-2}$)] & 4.84 $\pm$ 0.16 $^{(b)}$& 4.30 $\pm$ 0.12 $^{(f)}$& 4.361$^{+0.007}_{-0.008}$ $^{(i)}$& 4.15 $\pm$ 0.09 $^{(n)}$ & 4.1 $\pm$ 0.3 $^{(p)}$\\
 $v$ \rm{sin} $i_*$ [km s$^{-1}$] &0.330$^{+0.091}_{-0.066}$ $^{(a)}$& 3.5 $\pm$ 1.0 $^{(g)}$& 4.49 $\pm$ 0.50 $^{(g)}$& 71.4 $\pm$ 0.2 $^{(n)}$ & 111.40 $\pm$ 1.27 $^{(q)}$\\
 \\
 \underline{Planet Parameters} \\
        \textit{P} [days] & 2.64389803$^{+0.00000026}_{-0.00000026}$ $^{(a)}$ & 2.218577$^{+0.000009}_{-0.000010}$ $^{(e)}$ & 3.52474859$^{+0.00000038}_{-0.00000038}$ $^{(g)}$ & 2.73476550$^{+0.00000033}_{-0.00000033}$ $^{(n)}$ & 1.48111871$^{+0.00000016}_{-0.00000016}$ $^{(r)}$\\
        $T_0$ $[\rm BJD_{TDB}]$ & 2454865.084034$^{+0.000035}_{-0.000035}$ $^{(a)}$  &  2458334.990899$^{+0.000726}_{-0.000781}$ $^{(e)}$  & 2454560.80588$^{+0.00008}_{-0.00008}$ $^{(k)}$& 2458835.661885$^{+0.000073}_{-0.000073}$ $^{(n)}$ & 2458415.362562$^{+0.000081}_{-0.000081}$ $^{(r)}$ \\
    $T\rm_{14}$ $[\rm h]$ & 1.009 $\pm$ 0.034 $^{(c)}$ & 1.84 $\pm$ 0.04 $^{(e)}$ & 2.978 $\pm$ 0.051 $^{(l)}$ & 3.5112$^{+0.0233}_{-0.0221}$ $^{(o)}$ & 3.9158 $\pm$ 0.0115 $^{(q)}$\\
    $i$ [deg] & 86.858$^{+0.049}_{-0.052}$ $^{(a)}$ & 85.690$^{+0.095}_{-0.097}$ $^{(e)}$ & 86.710 $\pm$ 0.050 $^{(g)}$& 83.51 $\pm$ 0.09 $^{(n)}$ & 86.79 $\pm$ 0.25 $^{(q)}$\\
    $a/R_\star$ & 14.54 $\pm$ 0.14 $^{(a)}$ & 8.73 $\pm$ 0.15 $^{(e)}$ & 8.76 $\pm$ 0.04 $^{(i)}$& 5.452 $\pm$ 0.028 $^{(n)}$ & 2.996 $\pm$ 0.069 $^{(p)}$\\
        $M_\mathrm{p}$ $[\rm M_{J}]$ & 0.0799$^{+0.0066}_{-0.0063}$ $^{(a)}$& 1.166$^{+0.052}_{-0.049}$ $^{(e)}$ & 0.682$^{+0.015}_{-0.014}$ $^{(i)}$ & 1.28 $\pm$ 0.17 $^{(n)}$& 2.88 $\pm$ 0.35 $^{(p)}$\\
    $R_\mathrm{p}/R_\star$ & 0.08310 $\pm$ 0.00027 $^{(d)}$& 0.1504$^{+0.0038}_{-0.0039}$ $^{(e)}$ & 0.12086 $\pm$ 0.00010 $^{(i)}$& 0.08957 $\pm$ 0.00012 $^{(n)}$& 0.08228 $\pm$ 0.00281 $^{(p)}$\\
        \textit{b} &  0.7972$^{+0.0053}_{-0.0055}$ $^{(c)}$& 0.656$^{+0.014}_{-0.015}$ $^{(e)}$& 0.499 $\pm$ 0.008 $^{(k)}$& 0.597$^{+0.022}_{-0.025}$ $^{(o)}$&  0.177 $\pm$ 0.014 $^{(q)}$\\
    $\lambda$ $[\rm deg]$ & 72$^{+33}_{-24}$ $^{(a)}$& -0.4 $\pm$ 0.2 $^{(h)}$& -1.6 $\pm$ 0.3 $^{(m)}$& -10.55 $\pm$ 0.27 $^{(n)}$& -85.78 $\pm$ 0.46 $^{(p)}$\\
        \hline         
        \end{tabular}
 \end{spacing}{}
\tablefoot{
        \tablefoottext{a}{\citet{Bourrier_2018};} \tablefoottext{b}{\citet{sweetcat_2014};} \tablefoottext{c}{\citet{Lanotte_2014};} \tablefoottext{d}{\citet{Turner_2016};}
        \tablefoottext{e}{\citet{Addison_2019};}
        \tablefoottext{f}{\citet{Sousa_2021};} \tablefoottext{g}{\citet{Bonomo_2017};} \tablefoottext{h}{\citet{Cegla_2016};} \tablefoottext{i}{\citet{Torres_2008};}\tablefoottext{j}{\citet{Sousa_2008};} \tablefoottext{k}{\citet{Evans_2015};}
        \tablefoottext{l}{\citet{Richardson_2006};} 
        \tablefoottext{m}{\citet{CB_2020};}\tablefoottext{n}{\citet{Tabernero_Kelt7};} 
        \tablefoottext{o}{\citet{Bieryla_2015};} \tablefoottext{p}{\citet{Borsa_2019};} \tablefoottext{q}{\citet{Gaudi_2017};} \tablefoottext{r}{\citet{Ivshina-Winn}.}
        }
\end{table*}

\begin{table*}[h]
 \caption{\label{tab:pams2}KELT-20, WASP-69, WASP-76, WASP-80, and WASP-127  system parameters adopted in this work.}
 \centering
 \begin{spacing}{1.4}
 \scriptsize
        \begin{tabular}{l c c c c c}
                \hline \hline
   & KELT-20 b & WASP-69 b & WASP-76 b & WASP-80 b & WASP-127 b\\
  \hline
  \underline{Stellar Parameters} \\
        $M_\star$ [M$_\sun$] &  1.89$^{+0.06}_{-0.05}$ $^{(a)}$& 0.826 $\pm$ 0.029 $^{(d)}$ & 1.458 $\pm$ 0.021 $^{(h)}$& 0.577$^{+0.051}_{-0.054}$ $^{(i)}$& 0.950 $\pm$ 0.020 $^{(j)}$\\
        $R_\star$ [R$_\sun$] & 1.60 $\pm$ 0.06 $^{(a)}$ & 0.813 $\pm$ 0.028 $^{(d)}$& 1.756 $\pm$ 0.071 $^{(h)}$ & 0.586$^{+0.017}_{-0.018}$ $^{(i)}$ & 1.333 $\pm$ 0.027 $^{(j)}$\\
        $T\rm_{eff}$ [K] & 8980$^{+90}_{-130}$ $^{(a)}$& 4715 $\pm$ 50 $^{(d)}$& 6329 $\pm$ 65 $^{(h)}$& 4143$^{+92}_{-94}$  $^{(i)}$& 5842 $\pm$ 14 $^{(k)}$\\
        $[\rm Fe/H]$ [dex] &  -0.02 $\pm$ 0.07 $^{(a)}$& 0.144 $\pm$ 0.077 $^{(d)}$ & 0.366 $\pm$ 0.053 $^{(h)}$& -0.13$^{+0.15}_{-0.17}$ $^{(i)}$ & -0.19 $\pm$ 0.01 $^{(k)}$\\
    $\log g$ [$\log_{10}$(cm s$^{-2}$)] & 4.31 $\pm$ 0.02 $^{(a)}$ & 4.535 $\pm$ 0.023 $^{(d)}$& 4.196 $\pm$ 0.106 $^{(h)}$ & 4.663$^{+0.015}_{-0.016}$ $^{(i)}$ & 4.23 $\pm$ 0.02 $^{(k)}$\\
 $v$ \rm{sin} $i_*$ [km s$^{-1}$] & 114 $\pm$ 3 $^{(a)}$& 2.2 $\pm$ 0.4 $^{(d)}$& 1.48 $\pm$ 0.28 $^{(h)}$& 1.27$^{+0.14}_{-0.17}$ $^{(i)}$ & 0.53$^{+0.07}_{-0.05}$ $^{(k)}$ \\
 \\
 \underline{Planet Parameters} \\
        \textit{P} [days] & 3.47410196$^{+0.00000106}_{-0.00000106}$ $^{(b)}$ & 3.8681382$^{+0.0000017}_{-0.0000017}$ $^{(d)}$ & 1.80988198$^{+0.00000064}_{-0.0000056}$ $^{(h)}$ & 3.06785234$^{+0.00000083}_{-0.00000079}$ $^{(i)}$ & 4.17806203$^{+0.00000088}_{-0.00000053}$ $^{(j)}$\\
        $T_0$ $[\rm BJD_{TDB}]$ & 2457503.120120$^{+0.00018}_{-0.00018}$ $^{(b)}$  &  2455748.83344$^{+0.00018}_{-0.00018}$ $^{(d)}$  & 2458080.626165$^{+0.000418}_{-0.000367}$ $^{(h)}$& 2456487.425006$^{+0.000023}_{-0.000025}$ $^{(i)}$ & 2456776.62124$^{+0.00023}_{-0.00028}$ $^{(j)}$\\
    $T\rm_{14}$ $[\rm h]$ & 3.5755$^{+0.0218}_{-0.0211}$ $^{(c)}$ & 2.2296 $\pm$ 0.0288 $^{(d)}$ & 3.83 $^{(h)}$ & 2.1310$^{+0.0031}_{-0.0034}$ $^{(i)}$ & 4.3529$^{+0.0084}_{-0.0139}$ $^{(j)}$\\
    $i$ [deg] & 85.61 $\pm$ 0.19 $^{(b)}$ & 86.71 $\pm$ 0.20 $^{(d)}$ & 89.623$^{+0.005}_{-0.034}$ $^{(h)}$& 89.02$^{+0.11}_{-0.10}$ $^{(i)}$ & 87.84$^{+0.36}_{-0.33}$ $^{(j)}$\\
    $a/R_\star$ & 7.42 $\pm$ 0.13 $^{(c)}$ & 12.00 $\pm$ 0.46 $^{(e)}$ &4.08$^{+0.02}_{-0.06}$ $^{(h)}$& 12.63$^{+0.08}_{-0.13}$ $^{(i)}$ & 7.81$^{+0.11}_{-0.09}$ $^{(j)}$ \\
        $M_\mathrm{p}$ $[\rm M_{J}]$ & < 3.382 $^{(c)}$& 0.260 $\pm$ 0.017 $^{(d)}$ & 0.894$^{+0.014}_{-0.013}$ $^{(h)}$ & 0.538$^{+0.035}_{-0.036}$ $^{(i)}$& 0.1647$^{+0.0214}_{-0.0172}$ $^{(j)}$ \\
    $R_\mathrm{p}/R_\star$ & 0.11440$^{+0.00062}_{-0.00061}$ $^{(c)}$& 0.134 $\pm$ 0.008 $^{(d)}$ & 0.10852$^{+0.00096}_{-0.00072}$  $^{(h)}$& 0.17137$^{+0.00037}_{-0.00039}$ $^{(i)}$& 0.10103$^{+0.00026}_{-0.00047}$ $^{(j)}$ \\
        \textit{b} &  0.503$^{+0.025}_{-0.028}$ $^{(c)}$& 0.686 $\pm$ 0.023 $^{(d)}$& 0.027$^{+0.13}_{-0.023}$ $^{(h)}$& 0.215$^{+0.020}_{-0.022}$ $^{(i)}$ & 0.29 $\pm$ 0.04 $^{(j)}$ \\
    $\lambda$ $[\rm deg]$ & 0.6 $\pm$ 4 $^{(a)}$& 0.4$^{+2.0}_{-1.9}$ $^{(g)}$& 61.28$^{+7.61}_{-5.06}$  $^{(h)}$& -14$^{+15}_{-14}$ $^{(i)}$& -128.41$^{+5.60}_{-5.46}$ $^{(k)}$\\
        \hline       
        \end{tabular}
 \end{spacing}{}
 \tablefoot{
        \tablefoottext{a}{\citet{Talens_2018};}  \tablefoottext{b}{\citet{CB_2019};}  \tablefoottext{c}{\citet{Lund_2017};}  \tablefoottext{d}{\citet{Anderson_2014};}  \tablefoottext{e}{\citet{Stassun_2017};}  \tablefoottext{f}{\citet{Sousa_2018};}  \tablefoottext{g}{\citet{CB_2017};}  \tablefoottext{h}{\citet{Ehrenreich_2020};}   \tablefoottext{i}{\citet{Triaud_2015};}  \tablefoottext{j}{\citet{Seidel_127};}  \tablefoottext{k}{\citet{Allart_2020}.}
        }
\end{table*}
\pagebreak

\begin{figure*}
    \centering
    \includegraphics[width=0.85\textwidth]{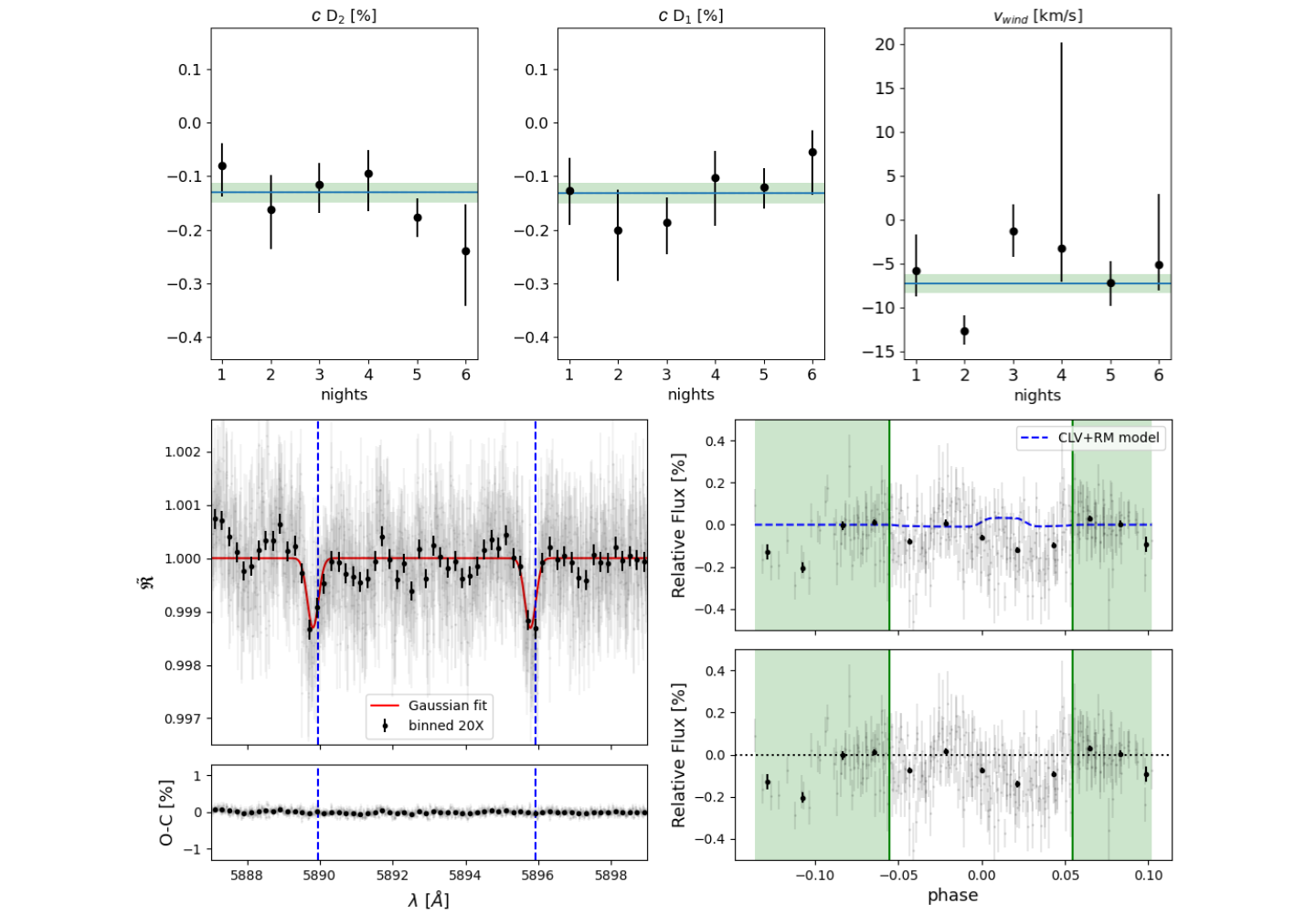}
    \caption{Same as Fig. \ref{fig:HD189733} but for KELT-9 b. For this target, we did not apply a selection since all nights present a detection on both D lines. For this reason, the 
    upper panels show only one average value and it is indicated by the green band. } 
    \label{fig:Kelt9}
\end{figure*}

\begin{figure*}
    \centering
    \includegraphics[width=0.85\textwidth]{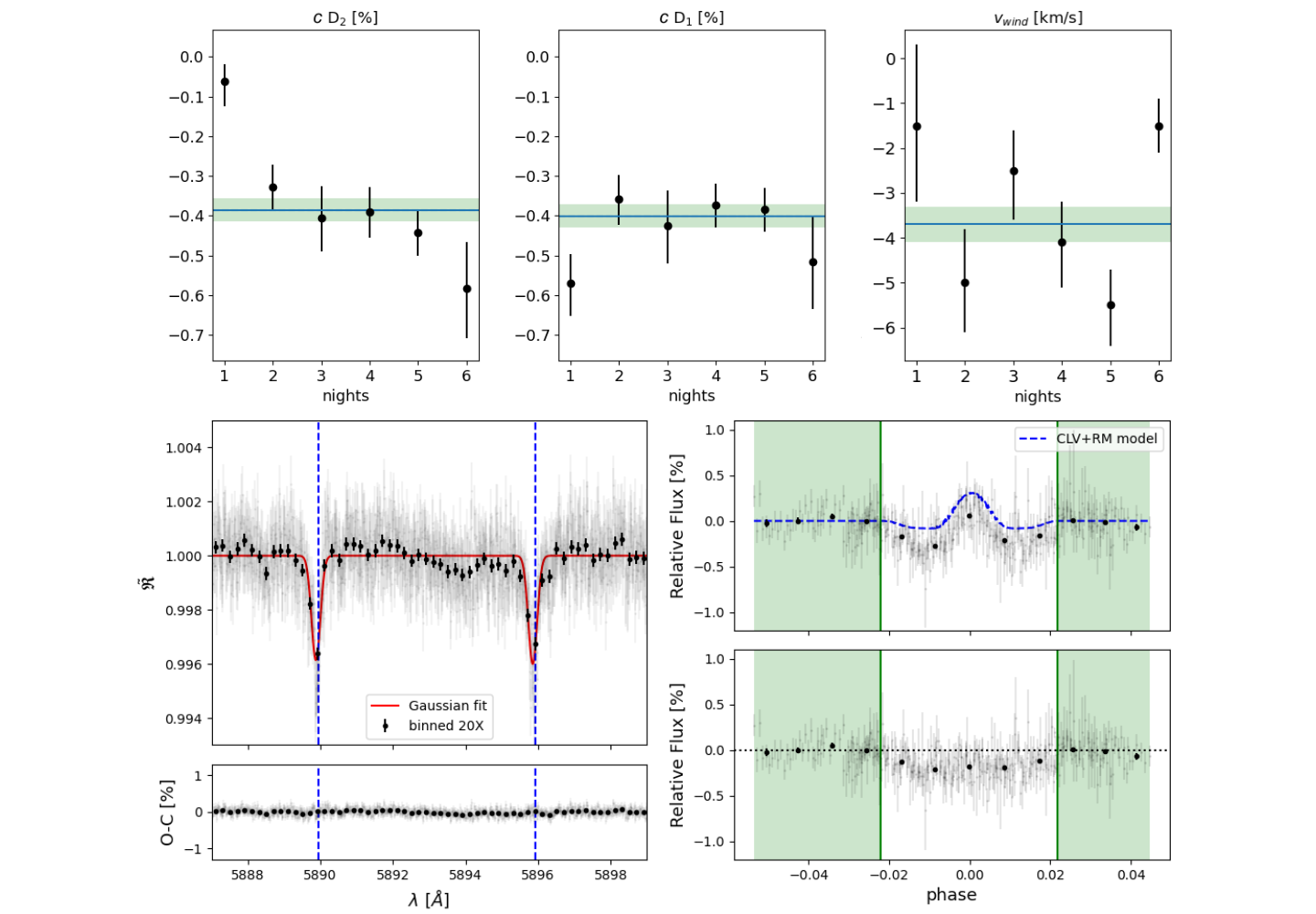}
    \caption{Same as Fig. \ref{fig:HD189733} but for KELT-20 b. For this target, we did not apply a selection since all nights present a detection on both D lines. For this reason, the 
    upper panels show only one average value and it is indicated by the green band.}
    \label{fig:Kelt20}
\end{figure*}

\begin{figure*}
    \centering
    \includegraphics[width=0.85\textwidth]{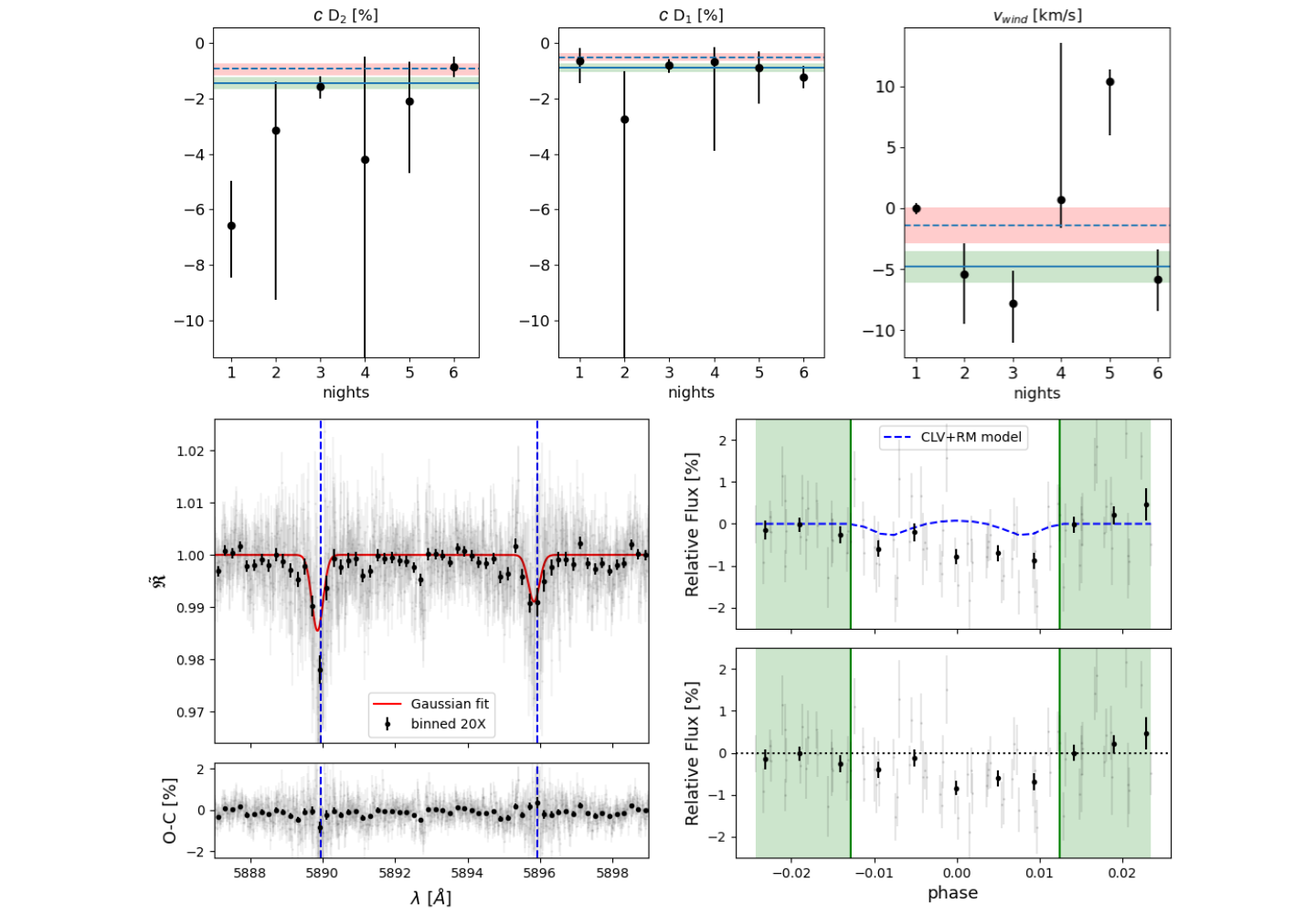}
    \caption{Same as Fig. \ref{fig:HD189733} but for WASP-69 b. The results are referred to the average spectrum after excluding W69-N4 and W69-N5.}  
    \label{fig:WASP69}
\end{figure*}

\begin{figure*}
    \centering
    \includegraphics[width=0.85\textwidth]{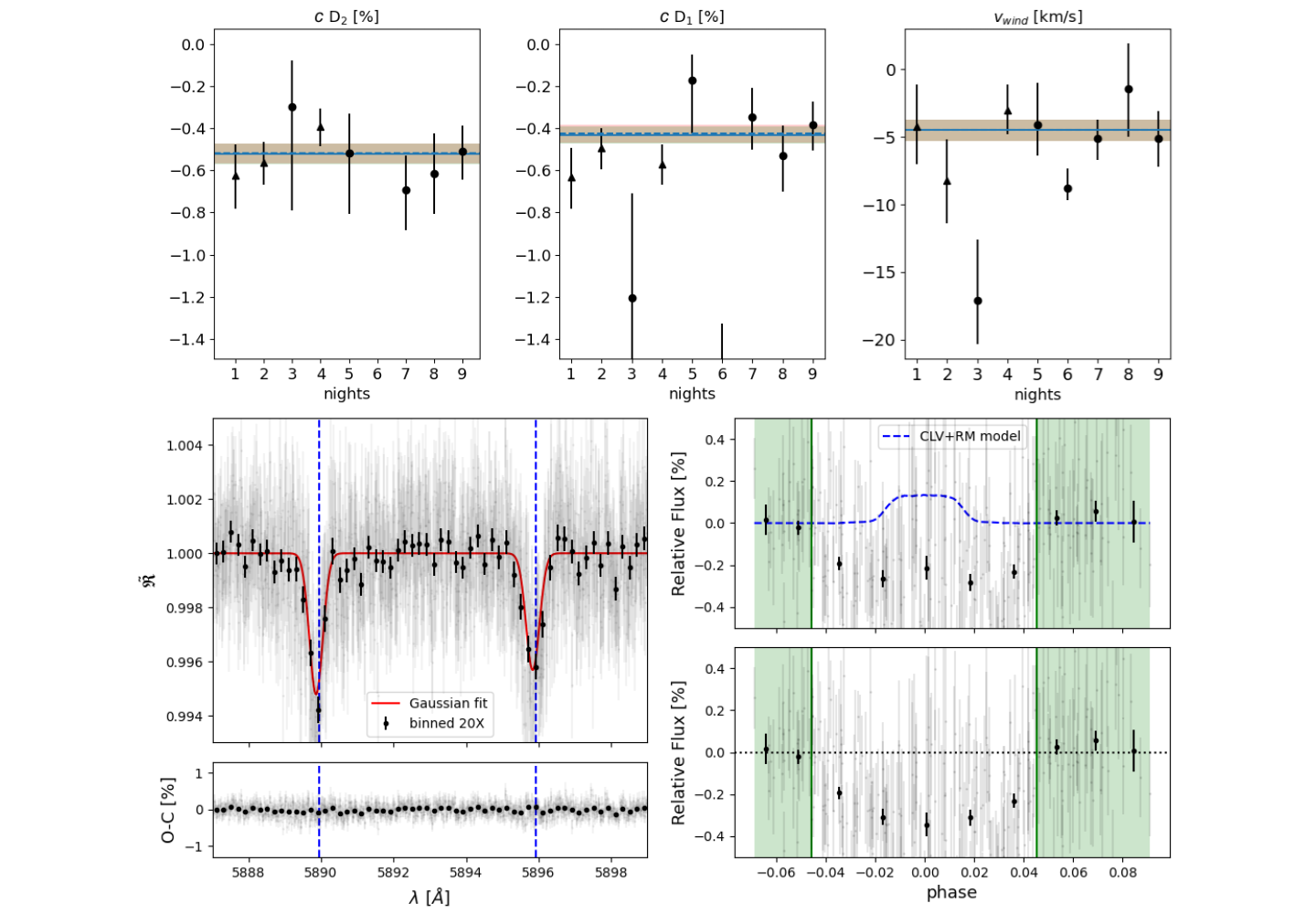}
    \caption{Same as Fig. \ref{fig:HD189733} but for WASP-76 b. The contrast values for both D lines in W76-N6 are out of scale. The results are referred to the average spectrum after excluding W76-N6.}
    \label{fig:WASP-76}
\end{figure*}

\begin{figure*}
    \centering
    \includegraphics[width=0.85\textwidth]{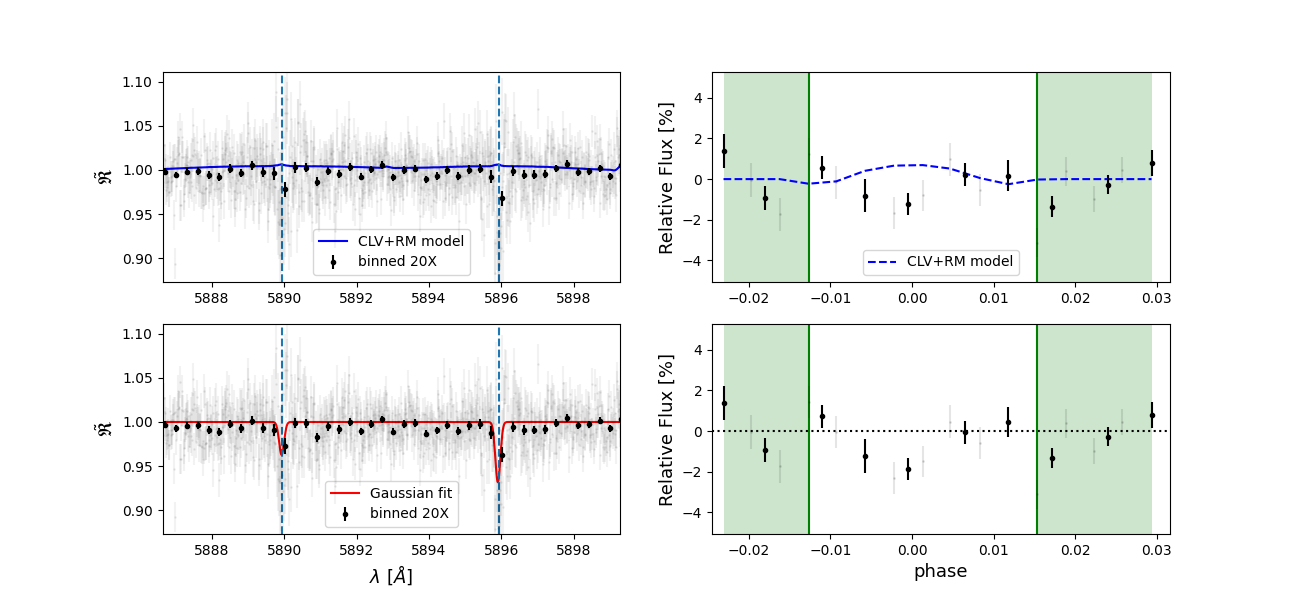}
    \caption{Same as Fig. \ref{fig:HD209458} but for WASP-80 b. The plots refer to the results obtained considering only W80-N6.}
    \label{fig:WASP-80}
\end{figure*}

\begin{figure*}
    \centering
    \includegraphics[width=0.85\textwidth]{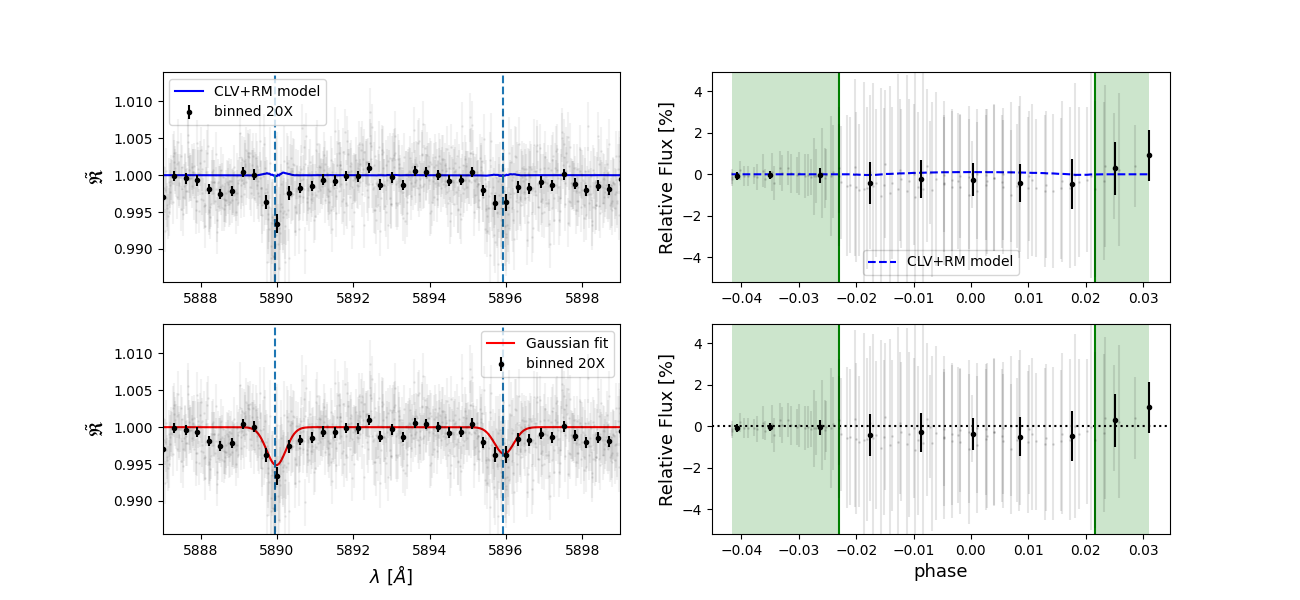}
    \caption{Same as Fig. \ref{fig:HD209458} but for WASP-127 b. The plots refer to the results obtained after excluding W127-N3, W127-N4 and W127-N5.}
    \label{fig:WASP127}
\end{figure*}

\end{document}